\documentclass{article}

\usepackage{arxiv}

\usepackage[utf8]{inputenc} % allow utf-8 input
\usepackage[T1]{fontenc}    % use 8-bit T1 fonts
\usepackage{hyperref}       % hyperlinks
\usepackage{url}            % simple URL typesetting
\usepackage{booktabs}       % professional-quality tables
\usepackage{amsfonts}       % blackboard math symbols
\usepackage{amssymb,amsthm,amsmath}
\usepackage{nicefrac}       % compact symbols for 1/2, etc.
\usepackage{microtype}      % microtypography
\usepackage{lipsum}         % dummy text
\usepackage{graphicx}       % images
\graphicspath{ {./images/} }
\usepackage{algorithm2e}    % algorithm environment
\usepackage{xcolor}
\usepackage{makecell}

\hypersetup{
    colorlinks=true,
    linkcolor=blue,
    filecolor=magenta,
    urlcolor=cyan,
}

\title{Differentiable Causal Discovery of Linear Non-Gaussian Acyclic Models Under Unmeasured Confounding}

\author{
  Yoshimitsu Morinishi \\
  Graduate School of Data Science \\
  Shiga University; Accenture Co. Ltd. \\
  \texttt{s7024104@st.shiga-u.ac.jp} \\
  \And
  Shohei Shimizu \\
  Faculty of Data Science \\
  Shiga University; RIKEN Center for Advanced Intelligence Project \\
  \texttt{shohei-shimizu@biwako.shiga-u.ac.jp} \\
}

\begin{document}
\maketitle

\begin{abstract}
We propose a novel score-based causal discovery method, named ABIC LiNGAM, which extends the linear non-Gaussian acyclic model (LiNGAM) framework to address the challenges of causal structure estimation in scenarios involving unmeasured confounders. By introducing the assumption that error terms follow a multivariate generalized normal distribution, our method leverages continuous optimization techniques to recover acyclic directed mixed graphs (ADMGs), including causal directions rather than just equivalence classes. We provide theoretical guarantees on the identifiability of causal parameters and demonstrate the effectiveness of our approach through extensive simulations and applications to real-world datasets.
\end{abstract}

% keywords can be removed
% \keywords{Causal discovery \and LiNGAM \and Unmeasured confounding \and Multivariate generalized normal distribution \and Continuous optimization}

\section{Introduction}
Uncovering the causal relationships among factors is a fundamental objective in many domains. Although randomized controlled trials (RCTs) are the most effective method for identifying causal relationships, their implementation is often infeasible due to high costs or ethical concerns. Consequently, it is crucial to develop methods for estimating causal relationships from observational data, when experimental studies are not feasible.

Many algorithms have been developed to uncover causal structures assuming the absence of unmeasured variables (\cite{spirtes2000causation}, \cite{chickering2002optimal}, \cite{shimizu2006linear}, among others). This assumption implies that all information relevant to the observed phenomena is fully observed and retained. For instance, the linear non-Gaussian acyclic model (LiNGAM) (\cite{shimizu2006linear}) demonstrates that causal structures can be identified under the assumptions of continuous variables, linear functional forms, non-Gaussian distributions for exogenous variables, the independence of error terms (i.e., no unmeasured variables), and acyclicity.

However, the assumption of no unmeasured variables is restrictive. The absence of unmeasured variables is often unrealistic in practical applications. For example, when analyzing factors influencing customer purchasing behavior, factors such as income, occupation, and address can significantly impact behavior but are challenging to capture. When this assumption is violated, it becomes difficult to identify causal structures using methods that assume no unmeasured variables. Therefore, relaxing this assumption is critical for broadly applying causal discovery frameworks.

The issue of unmeasured variables can be addressed from the perspective of graph representations of variable relationships. The use of directed acyclic graphs (DAGs) is a standard approach for modeling causal relationships among observed variables; however, they cannot account for the influence of unmeasured variables. To overcome this limitation, acyclic directed mixed graphs (ADMGs) were introduced. ADMGs include both directed and bidirectional edges, enabling the representation of latent variable-induced covariation and confounding effects, thus capturing more complex causal structures and constraints beyond what DAGs can handle. This capability also allows for the handling of nonparametric equality constraints such as Verma constraints (\cite{Vermaconstraint}), making ADMGs a more flexible and powerful tool for causal analysis.

\cite{maeda2020rcd} introduced repetitive causal discovery (RCD), which estimates causal structures in LiNGAM by iteratively testing the independence of residuals between observed variables using regression analysis. This method distinguishes between variable pairs influenced by unmeasured variables (represented by bidirectional arrows) and those not influenced (represented by unidirectional arrows), thereby estimating causal graphs while accounting for unmeasured variables.

\cite{BANG} proposed the bow-free acyclic non-Gaussian (BANG) method, which estimates causal structures by performing regression, calculating residuals, and iteratively conducting independence tests using higher-order moments of residuals. Assuming non-Gaussian data, this method consistently recovers bow-free ADMGs (referred to as bow-free acyclic path diagrams, or BAPs in their study), going beyond Markov equivalence classes to identify accurate causal structures.

\cite{bhattacharya2021differentiable} proposed ABIC, a method that frames causal discovery as a continuous optimization problem. Using differentiable constraints, ABIC estimates various types of graphs, such as ancestral ADMGs, arid ADMGs, and bow-free ADMGs, from observational data, recovering causal structures up to Markov equivalence classes for linear Gaussian structural equation models.

\cite{hoyer2008estimation} proposed a method for estimating causal directions among variables with unmeasured variables in LiNGAM. By explicitly incorporating unmeasured variables into the model and employing independent component analysis (ICA), this approach estimates causal directions and demonstrates its effectiveness in identifying causal structures in small datasets.

We build on the method proposed by \cite{bhattacharya2021differentiable} and introduce ABIC LiNGAM, a novel approach for estimating causal structures, including orientations, rather than just Markov equivalence classes, in the presence of unmeasured variables. Specifically, this method assumes that the error terms follow a common non-Gaussian distribution known as the multivariate generalized normal distribution and employs continuous optimization to recover ADMGs.

The contributions of this study are as follows: (1) it demonstrates that bow-free ADMGs can be recovered, rather than only Markov equivalence classes, when data are linear and follow a multivariate generalized normal distribution; (2) it shows that Markov equivalence classes can still be estimated when data follow a normal distribution, thereby providing a more general framework applicable to various distributions; and (3) it proves the identifiability of parameters in ADMGs under the assumption of multivariate generalized normality of error terms. Causal discovery methods can be broadly categorized into constraint-based approaches, which estimate graph structures based on statistical conditional independence tests of variables in the data-generating probability distribution, and score-based approaches, which maximize the scores (e.g., log-likelihood) of graphs given the data to estimate causal structures. The proposed method is the first score-based approach within the LiNGAM framework to estimate causal structures in the presence of unmeasured variables.

\section{Problem Definition}

\subsection{Representation by Linear SEM}

In this section, we review linear SEMs and their graphical representations. We use uppercase letters (e.g., $X$) to denote variables or nodes in the graph and indexed uppercase letters (e.g., $X_i$) to denote specific variables or nodes. We also use the following standard matrix notation: $A_{ij}$ denotes the element in the $i$th row and $j$th column of matrix $A$, $A_{-i,-j}$ denotes the submatrix of $A$ obtained by removing the $i$th row and $j$th column, and $A_{:,i}$ denotes the $i$th column of $A$.

Additionally, for each vertex $i$ belonging to set $V$, let $\{\mathrm{pa}(i) \mid i \in V\}$ and $\{\mathrm{sp}(i) \mid i \in V\}$ be two families of index sets. The vertex set of \( G \) is the index set \( V \), and \( G \) contains the edge \( j \to i \) if and only if \( j \in Pa(i) \) and the edge \( j \leftrightarrow i \) if and only if \( j \in Sp(i) \) (or equivalently, \( i \in Sp(j) \)). Furthermore, $\{\mathrm{sp}(i) \mid i \in V\}$ satisfies the following symmetry condition: for any $j \in V$, $j \in \mathrm{sp}(i)$ holds if and only if $i \in \mathrm{sp}(j)$. These two families of sets $\{\mathrm{pa}(i) \mid i \in V\}$ and $\{\mathrm{sp}(i) \mid i \in V\}$ define the system of structural equations.

\subsubsection{Linear SEM}

We consider linear SEMs for $d$ variables, parameterized by a weight matrix $\theta \in \mathbb{R}^{d \times d}$. For each variable $X_i \in X$, the structural equation is 
\begin{equation}
  X_i = \sum_{j \in \mathrm{pa}(i)} \theta_{ij} X_j + \epsilon_i, \quad i \in V
\end{equation}
Here, the noise terms $\epsilon_i$ are mutually independent. In this case, $\mathrm{sp}(i) = \emptyset$ for all $i$, since no unmeasured variables exist. The graph $G$ and corresponding binary adjacency matrix $D \in \{0,1\}^{d \times d}$ are defined as follows: An edge $X_j \rightarrow X_i$ exists in $G$ if and only if $\theta_{ij} \neq 0$, in which case $D_{ij} = 1$. The graph $G$ is acyclic if and only if $\theta$ can be created as an upper triangular matrix by the permutation of vertex labeling.

\subsubsection{Linear SEM with unmeasured variables}

An observed set of variables is causally insufficient if there exist unmeasured variables that are the ancestors of two or more observed variables in the system. In a linear structural equation model (SEM), these unmeasured variables often manifest as dependencies among the error terms \cite{pearl2009causality}. Consider a $d$-dimensional random vector $X = (X_1, \dots, X_d)$ represented by real-valued matrices $\delta, \Omega \in \mathbb{R}^{d \times d}$. For each $X_i$, the structural equation takes the following form:
\begin{equation}
    \label{ADMGs linear model generalized}
    X_i = \sum_{j \in \mathrm{pa}(i)} \delta_{ij} X_j + \epsilon_i, \quad i \in V.
\end{equation}
Here, $\epsilon = (\epsilon_1,\dots,\epsilon_d)$ is a vector of error terms with zero mean without loss of generality and is not necessarily Gaussian. Allowing non-Gaussian noise terms accommodates a wider class of underlying distributions and may improve identifiability via higher-order moments or nonsymmetric distributional features \cite{shimizu2006linear,BANG}.

In the special case where a given variable $X_i$ has no unmeasured variables, its error term $\epsilon_i$ may be independent of all the others. However, if unmeasured variables influence multiple observed variables, their corresponding error terms become dependent on each other. These dependencies are captured by the matrix $\Omega = \mathbb{E}[\epsilon \epsilon^\top]$, which does not need to be diagonal. For the Gaussian noise, the marginalized distribution of $X$ is a zero-mean multivariate normal with a covariance matrix as follows:
\begin{equation}
    \Sigma = (I - \delta)^{-1} \Omega (I - \delta)^{-\top},
\end{equation}
And the same covariance structure can be considered for non-Gaussian errors, at least at the level of second moments. In a non-Gaussian setting, higher-order moments and distributional asymmetries can be exploited to identify causal directions and latent structures.

The induced graph $G$ is an ADMG that includes both directed ($\rightarrow$) and bidirected ($\leftrightarrow$) edges. The graph $G$ and associated binary adjacency matrices $D \in \{0,1\}^{d \times d}$ and $B \in \{0,1\}^{d \times d}$ are defined as follows: a directed edge $X_j \rightarrow X_i$ exists in $G$ if and only if $\delta_{ij} \neq 0$, in which case $D_{ij} = 1$. A bidirected edge $X_j \leftrightarrow X_i$ exists in $G$ if and only if $\Omega_{ij} \neq 0$ (symmetry ensures $\Omega_{ji} \neq 0$), in which case $B_{ij} = B_{ji} = 1$. In the special case where there are no unmeasured variables, the ADMG reduces to a DAG, and the $B$ matrix is a zero matrix.

In summary, this framework does not restrict the noise terms to be Gaussian, allowing a broader class of SEMs that can represent latent variable-induced dependencies through non-Gaussian distributions. By leveraging non-Gaussianity, one can potentially achieve stronger identifiability and more robust causal inferences than would be possible under Gaussian assumptions alone.

\subsection{Motivation Example}

As discussed in Section 2.1.1, when there are no unmeasured variables, the observed variables can be represented as a DAG. Thus, the problem reduces to estimating the structure of a DAG. However, when unmeasured variables are present, a DAG cannot adequately represent the relationships between variables while accounting for such unmeasured variables. Therefore, we use an  ADMG, which can represent latent variable-induced covariation and confounding through directed and bidirected edges. Consequently, in the presence of unmeasured variables, the problem reduces to estimating the structure of an ADMG. This section builds on the work of \cite{bhattacharya2021differentiable}.

\begin{figure}[!h]
    \vspace{8em}
  \centering
  \includegraphics[bb=0 0 1100 70, width=\textwidth]{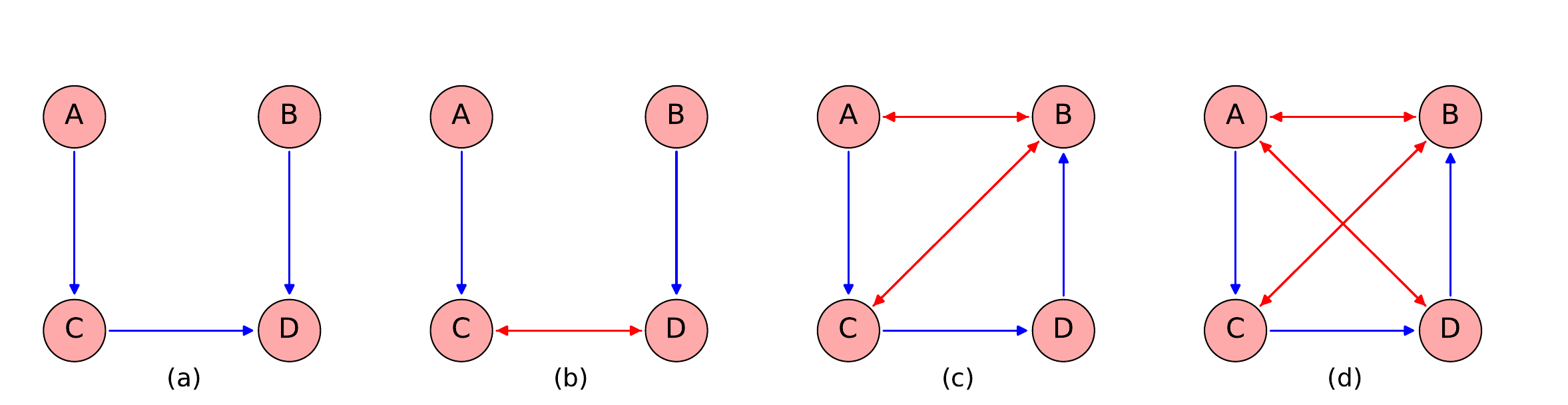}
  \caption{(a) DAG without unmeasured variables. (b) Ancestral ADMGs. (c) Arid ADMGs. (d) bow-free ADMGs.}
  \label{fig:causal-graphs}
\end{figure}

 % 色変える！
 Figure 1(a) depicts a DAG, which represents the relationships among the variables in the absence of unmeasured variables. Figures (b), (c), and (d) illustrate examples of ADMGs that we aim to estimate in this study. Figure (b) shows an ancestral ADMG, where no directed path \( X_i \to \cdots \to X_j \) and bidirected edges \( X_i \leftrightarrow X_j \) exist simultaneously in \( G \) for any pair of vertices \( X_i, X_j \in X \). Figure (c) shows an arid ADMG that does not contain any c-trees. A c-tree is a subgraph of \( G \) where its directed edges form a directed tree, and its bidirected edges form a single bidirectional connected component within the subgraph. For details on c-trees, see \cite{shpitser2006identification}. Figure (d) shows a bow-free ADMGs, where no pair of vertices \( X_i \to X_j \) and \( X_i \leftrightarrow X_j \) both exist in \( G \). These three types of ADMGs exhibit an inclusion relationship, with bow-free ADMGs being the most general type of ADMGs.

\[
\text{Ancestral} \subset \text{Arid} \subset \text{Bow-free}
\]

In \cite{bhattacharya2021differentiable}, these ADMGs are expressed as differentiable constraints, allowing the selection of the appropriate ADMGs type to be estimated based on the data. This study adopts the differentiable constraints proposed by \cite{bhattacharya2021differentiable}, enabling the selection of suitable ADMGs types according to the data.

\subsection{Identifiability in the Model}

This section discusses the identifiability of the parameters in bow-free ADMGs where the error terms follow a multivariate generalized normal distribution. Brito and Pearl (2002) proved that given a bow-free ADMG model, the parameters are almost everywhere identifiable from the observed covariance matrix. Since this fact is often utilized under the assumption that the error terms are Gaussian, we show in this study that it also applies to bow-free ADMG models when the error terms follow a multivariate generalized normal distribution. Furthermore, we discuss \cite{BANG}, who demonstrated that the model can identify causal directions, and not only Markov equivalence classes, using the non-Gaussianity of error terms. This study also provides evidence that causal directions can be estimated.

\subsubsection{On the Identifiability of Parameters in bow-Free ADMGs with Multivariate Generalized Normal Distributions}

\cite{brito2002recursive} demonstrated that a bow-free ADMG model is almost always identifiable from the observed covariance matrix. 
As the argument in \cite{brito2002recursive} primarily assumes that the error terms are Gaussian, we follow \cite{brito2002recursive} to demonstrate that when the error terms have a multivariate generalized normal distribution, a bow-free ADMG model is almost always identifiable from the observed covariance matrix. (See the Appendix for the proof.)

\textbf{Theorem 1}

Let $G$ be a bow-free ADMGs with error terms following a multivariate generalized normal distribution, and let the set of parameters of $G$ be $\theta = \{\delta, \Omega\}$. Then, for almost all $\theta$, the following holds:
\[
\Sigma(\theta) = \Sigma(\theta')
\]
implies $\theta = \theta'$.

In other words, if two parameter sets $\theta$ and $\theta'$ yield the same covariance matrix $\Sigma$, then $\theta$ and $\theta'$ must be identical, except possibly when $\theta$ belongs to a set of Lebesgue measure zero.

\subsubsection{Identifiability of the structure in bow-Free ADMGs Using Non-Gaussianity}

\cite{BANG} demonstrated that when the data are non-Gaussian and correspond to a bow-free ADMG, the bow-free ADMG can be consistently recovered, including both the Markov equivalence class and causal directions. The multivariate generalized normal distribution assumed for the observed data in this study is non-Gaussian except in the special case. Therefore, unlike \cite{bhattacharya2021differentiable}, who were limited to recovering the Markov equivalence class, it is expected that the causal directions can also be estimated.

\section{Decomposition of the Log-Likelihood Function of the Multivariate Generalized Normal Distribution}

\subsection{Probability Density Function of the Multivariate Generalized Normal Distribution}

As defined by \cite{gomez1998power}, the probability density function of the multivariate generalized Gaussian distribution (MGGD) is given by:
\begin{equation}
    \label{mggd density function}
    f(X \mid \mu, \Sigma, \beta) 
    = \frac{\Gamma\!\bigl(\tfrac{p}{2}\bigr)}{\pi^{\tfrac{p}{2}} \,\Gamma\!\bigl(\tfrac{p}{2\beta}\bigr)}
      \cdot \frac{\beta}{2^{\tfrac{p}{2\beta}} \,\lvert \Sigma \rvert^{\tfrac{1}{2}}}
      \exp\!\Bigl(-\tfrac{1}{2} \bigl((X - \mu)^\top \Sigma^{-1} (X - \mu)\bigr)^\beta\Bigr),
\end{equation}
where \(X\) is a \(p\)-dimensional random vector (\(p \ge 1\)) that follows a power-exponential distribution with parameters \(\mu\), \(\Sigma\), and \(\beta\). Specifically, \(\mu \in \mathbb{R}^p\), \(\Sigma\) is a \((p \times p)\) positive-definite symmetric matrix, and \(\beta \in (0, \infty)\). \(\Gamma(\cdot)\) denotes the gamma function. Notably, the MGGD reduces to a multivariate normal distribution when \(\beta = 1\). In this distribution, if $\Sigma$ is a diagonal matrix, then the correlation coefficients between the components become zero. However, because the multivariate generalized normal distribution belongs to the elliptical family, a zero correlation does not imply independence. Nonetheless, if the components are assumed to be generated independently, they can be considered truly independent rather than merely uncorrelated.

\cite{gomez1998power} show that the MGGD is invariant under affine transformations. More precisely, if \( f(X \mid \mu_X, \Sigma_X, \beta)\) is the probability density function, then for the affine transformation
\begin{equation}
    Y = CX + b,
\end{equation}
where \(C\) is a nonsingular matrix, \(b\) is a vector in \(\mathbb{R}^p\), and the transformed variable \(Y\) follows \( f(Y \mid C\mu_X + b,\; C\,\Sigma_X\,C^\top,\; \beta)\). This indicates that the family of distributions remains within the same class under any nonsingular linear transformation and translation.

This affine transformation property is particularly useful in linear structural equation models (especially (\ref{ADMGs linear model generalized})). In such models, the relationships among the observed variables and the covariance structure of the error terms are modeled, leading to the covariance matrix \(\Sigma\) of the observed variables in the form
\[
    \Sigma = (I - \delta)^{-1} \,\Omega\, (I - \delta)^{-\top}.
\]
Furthermore, as discussed in Section 3.2.1, one can estimate the parameters \(\delta\) and \(\Omega\) from the observed covariance matrix. In other words, if the data follow an MGGD, then in principle \(\delta\) and \(\Omega\) can be estimated from an observed covariance matrix of the form \(\Sigma = (I - \delta)^{-1}\,\Omega\,(I - \delta)^{-\top}\).

\subsection{Log-Likelihood Function}

Assuming that in the ADMG graph $G = (V, E)$ of the linear model (\ref{ADMGs linear model generalized}), $N$ observations are drawn, where all the components of $\mu$ are zero (the mean vector is zero). The reason for setting all the components of $\mu$ to zero is to prevent notational clutter without loss of generality. In this case, the log-likelihood function is given by (\ref{mggd density function}) as follows:
\begin{equation}
    \label{mggd log density}
\begin{aligned}
    \ell(\mu, \Sigma, \beta | X) &= N \log \Gamma\left(\frac{p}{2}\right) + N \log \beta - \frac{p}{2} N \log \pi \\
    &\quad - N \log \Gamma\left(\frac{p}{2\beta}\right) - \frac{p}{2\beta} N \log 2 - \frac{N}{2} \log |\Sigma| \\
    &\quad - \frac{1}{2} \sum_{l=1}^{N} \left( X^{(l)\top} \Sigma^{-1} X^{(l)} \right)^\beta
\end{aligned}
\end{equation}

\subsection{Decomposition of the Log-Likelihood Function}

The main component of the proposed algorithm is the decomposition of the log-likelihood function of the multivariate generalized normal distribution, inspired by \cite{dorton2009computing}, who decomposed the log-likelihood function of the multivariate normal distribution.

Let $X_i \in \mathbb{R}^N$ denote the $i$th row of the observation matrix $X$ and $X_{-i}=X_{V/{i}}$ be the $(V \setminus \{i\}) \times N$ submatrix of $X$. We adopt the abbreviated notation \(X_C\) to represent the \(C \times N\) submatrix of the \(D \times N\) matrix \(X\), where \(C \leq D\).

\textbf{Theorem 2}

Let \(i \in X\) be a variable node in the ADMG graph $G = (V, E)$ of the linear model (\ref{ADMGs linear model generalized}). Let \(\|x\|^2 = x^\top x\) and define $\Omega_{ii. -i}$ as the conditional variance of \(\varepsilon_i\) given \(\varepsilon_{-i}\) as follows:
\begin{equation}
    \label{omega block}
    \Omega_{ii. -i} = \Omega_{ii} - \Omega_{i,-i} \Omega_{-i,-i}^{-1} \Omega_{-i,i}
\end{equation}
Here, \(\Omega_{i,-i}\) is the row vector obtained by removing the $i$th element from the $i$th row, \(\Omega_{-i,i}\) is the column vector obtained by removing the $i$th element from the $i$th column, and \(\Omega_{-i,-i}\) denotes the submatrix obtained by removing the $i$th row and $i$th column from \(\Omega\). Additionally, let \(\Omega_{-i,-i}^{-1} = (\Omega_{-i,-i})^{-1}\). Then, the log-likelihood function \(\ell(B, \Omega, \beta)\) of the graph $G = (V, E)$ can be decomposed as
\begin{equation}
\begin{aligned}
    \ell(\mu, \Sigma, \beta | X)  =  \\
    &\quad -\frac{N}{2} \log \Omega_{ii. -i} - \frac{N}{2} \log \det(\Omega_{-i, -i}) \\
    &\quad -  \frac{1}{2} \sum_{l=1}^{N} \left( \Omega_{ii. -i}^{-1} \left( (X_i^{(l)} - \delta_{i,\mathrm{pa}(i)} X_{\mathrm{pa}(i)}^{(l)} - \Omega_{i,\mathrm{sp}(i)} \left( \Omega_{-i,-i}^{-1} \varepsilon_{-i}^{(l)} \right)_{\mathrm{sp}(i)} )^2 + \varepsilon_{-i}^{(l)\top} \Omega_{-i,-i}^{-1} \varepsilon_{-i}^{(l)} \right) \right)^\beta
\end{aligned}
\end{equation}

\textbf{Proof.}

By rearranging the log-likelihood function described in equation (\ref{mggd log density}), considering the constant parts unrelated to the coefficient matrices $\delta$ and $\Omega$, noting that the determinant $\det(I - \delta) = 1$ since $\delta$ is acyclic, and that $(I - \delta) X = \varepsilon$, and considering that the covariance matrix $\Sigma$ can be expressed using the adjacency matrix $B$ and $\Omega$ as $\Sigma = \operatorname{Var}(X) := (I - \delta)^{-1} \Omega (I - \delta)^{-\top}$, we obtain the following:
\begin{equation}
    \label{mggd organize log density}
    \begin{aligned}
        \ell(\mu, \Sigma, \beta | X) 
        % &= N \log \Gamma\left(\frac{p}{2}\right) + N \log \beta 
        % - \frac{p}{2} N \log \pi \\
        % &\quad - N \log \Gamma\left(\frac{p}{2\beta}\right) 
        % - \frac{p}{2\beta} N \log 2 
        % - \frac{N}{2} \log |\Sigma| \\
        % &\quad - \frac{1}{2} \sum_{l=1}^{N} \left( X^{(l)\top} \Sigma^{-1} X^{(l)} \right)^\beta \\
        &= - \frac{N}{2} \log |\Sigma| - \frac{1}{2} \sum_{l=1}^{N} \left( X^{(l)\top} \Sigma^{-1} X^{(l)} \right)^\beta \\
        &= \frac{N}{2} \log |(I - \delta)\Omega(I - \delta)^{\top}| 
        - \frac{1}{2} \sum_{l=1}^{N} \left( X^{(l)\top} (I - \delta)^\top \Omega^{-1} (I - \delta) X^{(l)} \right)^\beta \\
        &= -\frac{N}{2} \log |\Omega| 
        - \frac{1}{2} \sum_{l=1}^{N} \left( \varepsilon^{(l)\top} \Omega^{-1} \varepsilon^{(l)} \right)^\beta
    \end{aligned}
\end{equation}

We can partition \(\Omega\) as a block matrix:
\begin{equation}
    \label{Omega}
    \Omega = 
    \begin{pmatrix}
        \Omega_{ii} & \Omega_{i,-i} \\
        \Omega_{-i,i} & \Omega_{-i,-i}
    \end{pmatrix}
\end{equation}

Based on equations (\ref{omega block}) and (\ref{Omega}), we can rearrange $\log |\Omega|$ as shown in equation (\ref{log Omega}):
\begin{equation}
    \label{log Omega}
    \begin{aligned}
        \log |\Omega|
        &= \log \left( \Omega_{ii} - \Omega_{i,-i} \Omega_{-i,-i}^{-1} \Omega_{-i,i} \right) + \log |\Omega_{-i,-i}|  \\
        &= \log \Omega_{ii. -i} + \log |\Omega_{-i,-i}|
    \end{aligned}
\end{equation}

Next, we rearrange the term $\varepsilon^{(l)\top} \Omega^{-1} \varepsilon^{(l)}$ in $\frac{1}{2} \sum_{l=1}^{N} \left( \varepsilon^{(l)\top} \Omega^{-1} \varepsilon^{(l)} \right)^\beta$. We partition \(\Omega^{-1}\) as a block matrix:
\begin{equation}
    \label{Omega inverse}
    \Omega^{-1} = 
    \begin{pmatrix}
        \Omega_{ii} & \Omega_{i,-i} \\
        \Omega_{-i,i} & \Omega_{-i,-i}
    \end{pmatrix}^{-1} = \begin{pmatrix}
        \Omega_{ii. -i}^{-1} & -\Omega_{ii. -i}^{-1} \Omega_{i,-i} \Omega_{-i,-i}^{-1} \\
        -\Omega_{-i,-i}^{-1} \Omega_{-i,i} \Omega_{ii. -i}^{-1} & \Omega_{-i,-i}^{-1} + \Omega_{-i,-i}^{-1} \Omega_{-i,i} \Omega_{ii. -i}^{-1} \Omega_{i,-i} \Omega_{-i,-i}^{-1}
    \end{pmatrix}.
\end{equation}

Considering that \(\Omega^{-1}\) is a block matrix, we can rearrange $\varepsilon^{(l)\top} \Omega^{-1} \varepsilon^{(l)}$ as follows:
\begin{equation}
\begin{aligned}
    \varepsilon^{(l)\top} \Omega^{-1} \varepsilon^{(l)} &= 
    \begin{pmatrix}
        \varepsilon_{i}^{(l)} & \varepsilon_{-i}^{(l)\top} 
    \end{pmatrix}
    \begin{pmatrix}
        \Omega_{ii. -i}^{-1} & -\Omega_{ii. -i}^{-1} \Omega_{i,-i} \Omega_{-i,-i}^{-1} \\
        -\Omega_{-i,-i}^{-1} \Omega_{-i,i} \Omega_{ii. -i}^{-1} & \Omega_{-i,-i}^{-1} + \Omega_{-i,-i}^{-1} \Omega_{-i,i} \Omega_{ii. -i}^{-1} \Omega_{i,-i} \Omega_{-i,-i}^{-1}
    \end{pmatrix}
    \begin{pmatrix}
        \varepsilon_{i}^{(l)} \\
        \varepsilon_{-i}^{(l)}
    \end{pmatrix}  \\
    &=  \Omega_{ii. -i}^{-1} (\varepsilon_{i}^{(l)} - \Omega_{i,-i} \Omega_{-i,-i}^{-1} \varepsilon_{-i}^{(l)})^2 + \varepsilon_{-i}^{(l)\top} \Omega_{-i,-i}^{-1} \varepsilon_{-i}^{(l)}
\end{aligned}
\end{equation}

From this rearrangement, the log-likelihood function becomes equation (\ref{epsiron omega epsiron}):
\begin{equation}
    \label{epsiron omega epsiron}
    \ell(\mu, \Sigma, \beta | X)  = -\frac{N}{2} \log \Omega_{ii. -i} - \frac{N}{2} \log \det(\Omega_{-i, -i}) + \frac{1}{2} \sum_{l=1}^{N} \left( \Omega_{ii. -i}^{-1} \left( (\varepsilon_{i}^{(l)} - \Omega_{i,-i} \Omega_{-i,-i}^{-1} \varepsilon_{-i}^{(l)} )^2 + \varepsilon_{-i}^{(l)\top} \Omega_{-i,-i}^{-1} \varepsilon_{-i}^{(l)} \right) \right)^\beta
\end{equation}

By definition, the error term \(\varepsilon_i^{(l)} = X_i^{(l)} - \delta_{i,\mathrm{pa}(i)} X_{\mathrm{pa}(i)}^{(l)}\). Moreover, since we are dealing with bow-free ADMGs, we have $\Omega_{i,-i} \Omega_{-i,-i}^{-1} \varepsilon_{-i}^{(l)} = \Omega_{i,\mathrm{sp}(i)} \left( \Omega_{-i,-i}^{-1} \varepsilon_{-i}^{(l)} \right)_{\mathrm{sp}(i)}$. This yields the claimed decomposition.

\hfill \(\square\)

The decomposition of the log-likelihood function is based on decomposing the joint distribution of \(\varepsilon\) into the marginal distribution of \(\varepsilon_{-i}\) and conditional distribution \((\varepsilon_i \mid \varepsilon_{-i})\). In particular, as shown in (\ref{epsiron omega epsiron}), the squared term \((\varepsilon_{i}^{(l)} - \Omega_{i,-i} \Omega_{-i,-i}^{-1} \varepsilon_{-i}^{(l)} )^2\) represents the deviation of \(\varepsilon_i\) from its conditional expectation given \(\varepsilon_{-i}\), which plays a key role in deriving the likelihood decomposition. This idea leads to an approach similar to that of \cite{dorton2009computing}, who decomposed the log-likelihood function of the multivariate normal distribution and proposed an iterative algorithm. The steps of this algorithm are based on fixing the marginal distribution of \(\varepsilon_{-i}\) and estimating the conditional distribution. To fix the marginal distribution of \(\varepsilon_{-i}\), we must fix the submatrix of \(\Omega_{-i,-i}\) excluding the $i$th row and $i$th column and the submatrix of \(\delta_{-i,V}\) excluding the $i$th row. This is because \(\varepsilon_{-i}\) is determined depending on \(\Omega_{-i,-i}\) and \(\delta_{-i,V}\). When $\Omega_{-i,-i}$ and $\delta_{-i,V}$ are fixed, not only can $\epsilon_{-i}$ be computed but also the pseudo-variable is defined as
\begin{equation}
    \label{pseudo-variable-definition}
    Z_{-i} = \Omega^{-1} \epsilon_{-i}.
\end{equation}

From $Z_{-i} = \Omega_{-i,-i}^{-1} \varepsilon_{-i}$, it becomes clear that when \(\Omega_{-i,-i}\) and \(\delta_{-i,V}\) are fixed, the maximization of the log-likelihood function \(\ell(\delta, \Omega, \beta)\) can be solved by maximizing the following function to estimate $\delta$ and $\Omega$:
\begin{equation}
    \label{mggd organized decomposition log density2}
    \begin{aligned}
    \ell(\mu, \Sigma, \beta | X) \\  
    &= -\frac{N}{2} \log \Omega_{ii. -i} 
     -  \frac{1}{2} \sum_{l=1}^{N} \left( \Omega_{ii. -i}^{-1} \left( X_i^{(l)} - \delta_{i,\mathrm{pa}(i)} X_{\mathrm{pa}(i)}^{(l)} - \Omega_{i,\mathrm{sp}(i)} \left( \Omega_{-i,-i}^{-1} \varepsilon_{-i}^{(l)} \right)_{\mathrm{sp}(i)} \right)^2 \right)^\beta \\
     &= -\frac{N}{2} \log \Omega_{ii. -i} 
     -  \frac{1}{2\Omega_{ii. -i}^\beta}  \sum_{l=1}^{N} \left( \left( X_i^{(l)} - \sum_{j \in \mathrm{pa}(i)} \delta_{i,j} X_j^{(l)} - \sum_{k \in sp(i)} \Omega_{i,k} Z_k^{(l)}  \right)^2 \right)^\beta \\
    \end{aligned}
\end{equation}

Assuming $\beta \geq 1$, we rearrange equation (\ref{mggd organized decomposition log density2}) using Hölder's inequality(for details on the application of Hölder's inequality, see the APPENDIX).

\begin{equation}
    \label{mggd organized decomposition log density3}
    \begin{aligned}
    \ell(\mu, \Sigma, \beta | X) \\  
     &= -\frac{N}{2} \log \Omega_{ii. -i} 
     -  \frac{1}{2\Omega_{ii. -i}^\beta}  \sum_{l=1}^{N} \left( \left( X_i^{(l)} - \sum_{j \in \mathrm{pa}(i)} \delta_{i,j} X_j^{(l)} - \sum_{k \in sp(i)} \Omega_{i,k} Z_k^{(l)}  \right)^2 \right)^\beta \\
     &= -\frac{N}{2} \log \Omega_{ii. -i} 
     -  \frac{1}{2\Omega_{ii. -i}^\beta} \frac{N^{\beta-1}}{N^{\beta-1}} \sum_{l=1}^{N} \left( \left( X_i^{(l)} - \sum_{j \in \mathrm{pa}(i)} \delta_{i,j} X_j^{(l)} - \sum_{k \in sp(i)} \Omega_{i,k} Z_k^{(l)}  \right)^2 \right)^\beta \\
     &\geq -\frac{N}{2} \log \Omega_{ii. -i} -  \frac{1}{2\Omega_{ii. -i}^\beta} \frac{1}{N^{\beta - 1}} 
     \left( \sum_{l=1}^{N} \left( X_i^{(l)} - \sum_{j \in \mathrm{pa}(i)} \delta_{i,j} X_j^{(l)} - \sum_{k \in sp(i)} \Omega_{i,k} Z_k^{(l)}  \right)^2  \right)^\beta \\
     &= -\frac{N}{2} \log \Omega_{ii. -i} - \frac{1}{2\Omega_{ii. -i}^\beta} \frac{1}{N^{\beta - 1}} 
     \left\| X_i - \sum_{j \in \mathrm{pa}(i)} \delta_{i,j} X_j - \sum_{k \in sp(i)} \Omega_{i,k} Z_k  \right\|^{2\beta}
    \end{aligned}
\end{equation}

Maximizing equation (\ref{mggd organized decomposition log density3}) to estimate $\delta$ and $\Omega$ is equivalent to performing a regression of $X_i$ (as the target variable) on $X_j$ (the parent variable of $X_i$) and the pseudo-variables $Z_k$, considering the shape parameter $\beta$.

Utilizing these observations, in the next section, we propose a method for causal structure estimation using continuous optimization, considering the presence of unmeasured variables and assuming that the error variables follow a multivariate generalized normal distribution, using the decomposition results of the log-likelihood function organized in this section to estimate \(\hat{\delta}, \hat{\Omega}\).

\section{Proposed Method}

\subsection{Causal Discovery Based on Differentiable Scores}

Score-based methods aim to estimate causal structures by maximizing a graph's score (e.g., log-likelihood) given the data. Learning DAGs from data is an NP-hard problem because it is challenging to efficiently enforce combinatorial acyclicity constraints (\cite{chickering1996optimal}).

\cite{zheng2018dags} proposed a new approach for score-based DAG learning by converting the traditional combinatorial optimization problem (\ref{discrete optimization}) into a continuous optimization problem (\ref{continuous optimization}).

\begin{equation}
\label{discrete optimization}
\min_{\theta \in \Theta} F(\theta) \quad \text{subject to } G(\theta) \in \text{DAGs}
\end{equation}

\begin{equation}
\label{continuous optimization}
\min_{\theta \in \Theta} F(\theta) \quad \text{subject to } h(\theta) = 0,
\end{equation}

where $G(\theta)$ is a $d$-node graph induced by the weight matrix $\theta \in \mathbb{R}^{d \times d}$, and $F : \mathbb{R}^{d \times d} \to \mathbb{R}$ is a score function. $h : \mathbb{R}^{d \times d} \to \mathbb{R}$ is a smooth function over real matrices, and the constraint $h(\theta)=0$ can precisely characterize the acyclicity of the graph. Causal structure estimation via continuous optimization eliminates the need for specialized algorithms to explore the combinatorial space of DAGs and instead allows the use of standard numerical algorithms for constrained problems, making implementation particularly straightforward, as mentioned in \cite{zheng2018dags}.

The acyclicity constraint is defined as follows:

\begin{align}
    h(\mathbf{\theta}) = \text{trace}\left( e^{\mathbf{\theta} \circ \mathbf{\theta}} \right) - d = 0
\end{align}

Here, $\mathbf{\theta} \circ \mathbf{\theta}$ denotes the Hadamard product (element-wise multiplication), $\text{trace}\left( e^{\mathbf{\theta} \circ \mathbf{\theta}} \right)$ is the trace (sum of the diagonal elements) of the matrix exponential, and $d$ is the number of variables. This constraint ensures that the matrix $\mathbf{\theta}$ forms a DAG.

\cite{zheng2018dags} used the augmented Lagrangian method as a continuous optimization technique. This method solves constrained optimization problems using an objective function that includes penalty terms and is formulated as

\begin{align}
    \min_{\mathbf{\theta} \in \Theta} L(\mathbf{\theta}) + \lambda \|\mathbf{\theta}\|_1 + \alpha h(\mathbf{\theta}) + \frac{\rho}{2} h(\mathbf{\theta})^2 \quad \text{subject to} \quad h(\mathbf{\theta}) = 0.
\end{align}

Here, $\lambda$ is the weight of the regularization term, $\alpha$ is the Lagrange multiplier, and $\rho$ is the penalty coefficient.

\subsection{ABIC}

Although the differentiable score-based causal discovery method using continuous optimization proposed by \cite{zheng2018dags} has been successful in estimating causal structures in DAGs, it cannot be directly applied to ADMGs. This is because ADMGs require two adjacency matrices, $D$ and $B$, to represent the directed and bidirected edges, respectively. To extend the differentiable algebraic characterization to ADMGs, \cite{bhattacharya2021differentiable} proposed differentiable constraints for each causal graph. In \cite{bhattacharya2021differentiable}, three differentiable constraints (i.e., ancestral, arid, and bow-free)are proposed for ADMGs , as shown in Table 1.

\begin{table}[h]
    \label{tab:constraints}
    \centering
    \begin{tabular}{ll}
    \toprule
    \textbf{ADMGs} & \textbf{Algebraic Constraint} \\
    \midrule
    Ancestral & \(\text{trace}(e^D) - d + \text{sum}(e^D \circ B) = 0\) \\
    Arid & \(\text{trace}(e^D) - d + \text{Greenery}(D, B) = 0\) \\
    Bow-free & \(\text{trace}(e^D) - d + \text{sum}(D \circ B) = 0\) \\
    \bottomrule
    \end{tabular}
    \caption{Differentiable constraints for each causal graph in ADMGs (\cite{bhattacharya2021differentiable})}
\end{table}

For example, to estimate the causal structure of bow-free ADMGs, the constraint equation becomes (\ref{bow-free constraints}).

\begin{equation}
    \label{bow-free constraints}
h(\theta) = \text{tr}(e^D) - d + \text{sum}(D \circ B)
\end{equation}

Here, $\text{sum}(\cdot)$ denotes the sum of all the elements in a matrix. It has been proven that when $h(\theta) = 0$, the estimated graph corresponds to an ADMG (\cite{bhattacharya2021differentiable}). Essentially, $\text{tr}(e^D) - d$ signifies the standard acyclicity constraint for directed edges, and the latter term $\text{sum}(D \circ B)$ ensures that bidirected edges are not introduced when a directed edge exists (i.e., it enforces the bow-free ADMGs property).

\cite{bhattacharya2021differentiable} used the augmented Lagrangian method, similar to \cite{zheng2018dags}, to convert the problem into an optimization problem with a quadratic penalty term, and proposed ABIC, which solves the following primal equation at each iteration:

\begin{equation}
\min_{\theta \in \Theta} ABIC_{\lambda}(X; \theta) + \frac{\rho}{2} |h(\theta)|^2 + \alpha h(\theta),
\end{equation}

where $\rho$ is the weight of the penalty term, and $\alpha$ is the Lagrange multiplier. Then, the Lagrange multiplier is updated as $\alpha \leftarrow \alpha + \rho h(\theta^)$. Intuitively, by optimizing the primal equation with a large $\rho$, we force $h(\theta)$ to be very close to zero, thus satisfying the equality constraint.

\begin{algorithm}[H]
  \caption{ABIC}
  \KwIn{$X, \Omega, tol, max\_iterations, h, \rho, \alpha, \lambda$}
  \KwOut{Estimates $\delta^t$ and $\Omega^t$}
  
  Initialize the estimates $\delta^t$ and $\Omega^t$ and set $c = \ln(n)$\;
  Define $LS(\theta)$ as $\frac{1}{2n} \sum_{i=1}^d ||X_{\cdot,i} - X\delta_{\cdot,i} - Z^{(i)} \Omega_{\cdot,i}||^2$\;
  
  \For{$t = 1$ \KwTo $max\_iterations$}{
      \For{$i = 1$ \KwTo $d$}{
          Compute $\epsilon_i \leftarrow X_{\cdot,i} - X \delta^t_{\cdot,i}$\;
          Compute $Z^{(i)} \in \mathbb{R}^{n \times d}$ as $Z_{\cdot,i}^{(i)} = 0$ and $Z_{\cdot,-i}^{(i)} \leftarrow \epsilon_{-i} (\Omega^t_{-i,-i})^{-T}$\;
      }
      $\delta^{t+1}, \Omega^{t+1} \leftarrow \arg \min_{\theta \in \Theta} \{LS(\theta) + \frac{\rho}{2}|h(\theta)|^2 + \alpha h(\theta) + \lambda \sum_{i=1}^{\dim(\theta)} \tanh(c| \theta_i |) \}$\;
      \For{$i = 1$ \KwTo $d$}{
          Compute $\epsilon_i \leftarrow X_{\cdot,i} - X \delta^{t+1}_{\cdot,i}$\;
          Set $\Omega_{ii}^{t+1} \leftarrow \text{var}(\epsilon_i)$\;
      }
      \If{$|| \delta^{t+1} - \delta^t + \Omega^{t+1} - \Omega^t || < tol$}{
          \textbf{break}\;
      }
  }
  \textbf{return} $\delta^t, \Omega^t$\;
  
  \end{algorithm}

Inspired by \cite{dorton2009computing}, Bhattacharya et al. (2021) proposed a method for estimating the Markov equivalence class when the error terms are normally distributed within the framework of continuous optimization. Specifically, in the case of an ADMG model where the error terms are normally distributed, maximizing the likelihood corresponds to minimizing a least squares regression problem where each variable $i$ is regressed on its direct parents $X_j \rightarrow X_i$ and pseudo-variables $Z$ (formed from residual noise terms and the bidirected coefficients of their siblings). At each step, $Z$ is computed using the current parameter estimates, and the primal equation is solved. This procedure is repeated until convergence or a prespecified maximum number of iterations is reached. When the penalty is small (resulting in non-arid graphs), they expect that ABIC will not converge during the initial iterations of the augmented Lagrangian method. Therefore, they initially set the maximum number of iterations to be small and increase this number at each dual step. The penalty $\rho$ is increased according to a fixed schedule (multiplied by 10) to a maximum value of $10^{16}$ if the inequality in line 4 of algorithm1 is not satisfied. Simulations have shown that this method is effective in practice, and the convergence of algorithm1 within 10 to 15 steps of the augmented Lagrangian method is achieved (\cite{bhattacharya2021differentiable}).

\subsection{ABIC LiNGAM}

In this study, we extend the method proposed by \cite{bhattacharya2021differentiable} and present algorithm2 that can estimate causal structures when the error terms follow a multivariate generalized normal distribution. The basic framework is the same as that in \cite{bhattacharya2021differentiable}, but we consider that the error terms follow a multivariate generalized normal distribution and incorporate the shape parameter $\beta$ into the loss function. The shape parameter must be estimated from the observed data in advance. We use the estimated shape parameter $\hat{\beta}$. In addition, since the multivariate normal distribution corresponds to $\beta=1$, the proposed method can handle the normal distribution case, thereby generalizing the method of \cite{bhattacharya2021differentiable}. Depending on the data, it is possible to switch between non-Gaussianity and Gaussianity.

\begin{algorithm}[H]
  \caption{ABIC LiNGAM}
  \KwIn{$X, \Omega, tol, max\_iterations, h, \rho, \alpha, \lambda$}
  \KwOut{Estimates $\delta^t$ and $\Omega^t$}
  
  Initialize the estimates $\delta^t$ and $\Omega^t$ and set $c = \ln(n)$\;
  Define $LS(\theta)$ as $\frac{1}{2n} \sum_{i=1}^d ||X_{\cdot,i} - X\delta_{\cdot,i} - Z^{(i)} \Omega_{\cdot,i}||^{2\beta}$\;
  
  \For{$t = 1$ \KwTo $max\_iterations$}{
      \For{$i = 1$ \KwTo $d$}{
          Compute $\epsilon_i \leftarrow X_{\cdot,i} - X\delta^t_{\cdot,i}$\;
          Compute $Z^{(i)} \in \mathbb{R}^{n \times d}$ as $Z_{\cdot,i}^{(i)} = 0$ and $Z_{\cdot,-i}^{(i)} \leftarrow \epsilon_{-i}(\Omega^t_{-i,-i})^{-T}$\;
      }
      $\delta^{t+1}, \Omega^{t+1} \leftarrow \arg \min_{\theta \in \Theta} \{LS(\theta) + \frac{\rho}{2}|h(\theta)|^2 + \alpha h(\theta) + \lambda \sum_{i=1}^{\dim(\theta)} \tanh(c| \theta_i |) \}$\;
      \For{$i = 1$ \KwTo $d$}{
          Compute $\epsilon_i \leftarrow X_{\cdot,i} - X\delta^{t+1}_{\cdot,i}$\;
          Set $\Omega_{ii}^{t+1} \leftarrow \text{var}(\epsilon_i)$\;
      }
      \If{$|| \delta^{t+1} - \delta^t + \Omega^{t+1} - \Omega^t || < tol$}{
          \textbf{break}\;
      }
  }
  \textbf{return} $\delta^t, \Omega^t$\;
  
  \end{algorithm}

  \section{Experiments}

  \subsection{Simulation}
  
  In this study, we conducted simulations inspired by the experimental setup in \cite{bhattacharya2021differentiable}. We verified whether the proposed method could identify ADMGs, particularly the most common bowel-free causal structures. The data generation process was as follows:
  
  The causal relationships for each pair $(i, j)\ (i < j)$ were determined using a two-step process involving random sampling. First, a value was uniformly sampled from the range $[0, 1]$ to determine whether the pair $(i, j)$ was assigned a directional ($X_i \to X_j$) or a bidirectional relationship ($X_i \leftrightarrow X_j$). If the sampled value was below a predefined threshold for inclusion in the coefficient matrix, a directional relationship was assigned, and $\delta_{ij}$ was sampled uniformly from the interval $[-2.0, -0.5] \cup [0.5, 2.0]$, with the value placed in the coefficient matrix $\delta_{ij}$.
  
  A bidirectional relationship was assigned if the sampled value was within the range corresponding to the threshold of the adjacency matrix. In this case, a value was uniformly sampled from the interval $[-0.7, -0.4] \cup [0.4, 0.7]$ and symmetrically assigned to $\Omega_{ij}$ and $\Omega_{ji}$. No causal relationships were assigned to pairs that did not meet either threshold.
  
  The diagonal entries $\Omega_{ii}$ were determined separately. These values were sampled from the interval $\pm[0.4, 0.7]$. To ensure the positive definiteness of $\Omega$, an adjustment was made by adding $\sum(|\Omega_{i, -i}|)$ and an offset sampled from the interval $[0.1, 0.5]$. 
  
  The thresholds for inclusion in the coefficient and adjacency matrices controlled for the probability of assigning each type of relationship, and consequently, the number of relationships assigned. This probabilistic framework ensured that the structures of $\delta$ and $\Omega$ aligned with the desired causal model.

  In this study, to compare our proposed method, we considered bow-free ABIC (\cite{bhattacharya2021differentiable}), which was the basis of our research; FCI (\cite{spirtes2000causation}), a constraint-based method that can estimate causal structures in the presence of unmeasured variables; and BANG (\cite{BANG}), which can identify bow-free ADMGs (at the 5\% significance level), and compared their accuracies. In this experiment, since we tested on bow-free data, the most common in ADMGs, our proposed method, ABIC LiNGAM, was evaluated with bow-free constraints. We examined two patterns: one where the true value of the shape parameter $\beta$ of the multivariate generalized normal distribution was known, and one where it was unknown and needed to be estimated.
  
  The simulations were conducted under three patterns for sample size $n$ ($\lbrace 100, 500, 1000 \rbrace$), two patterns for the dimension $k$ of the observed variables ($\lbrace 5, 10 \rbrace$), and three patterns for the shape parameter $\beta$ of the multivariate generalized normal distribution ($\lbrace 1, 3, 5 \rbrace$) (when $\beta=1$, it became a multivariate normal distribution), resulting in $3 \times 2 \times 3 = 18$ scenarios. In this study, the number of simulations for each scenario was set to 50. The simulations were conducted using Python 3.8 and the NumPy and SciPy libraries.

  Using non-Gaussianity, the structure of bow-free ADMGs could be identified. We evaluated whether the direction could be estimated in this verification. Therefore, for ABIC, which originally assumes a linear Gaussian SCM and reconstructs up to the Markov equivalence class, we evaluated the output. Since FCI outputs a partial ancestral graph (PAG), we treated $A \circ\!\!\rightarrow B$, indicating that $B$ was not an ancestor of $A$, as $A \rightarrow B$, and treated $\circ\!\!-\!\!\circ$, indicating that there was no set that d-separated $A$ and $B$, as $A \leftrightarrow B$. Consequently, we evaluated the precision, recall, and F1-score for skeletons, arrowheads, and tails.
  
  \begin{figure}[htbp]
    \vspace{-10em}
      \centering
      \includegraphics[bb=0 0 2200 1600, width=\textwidth]{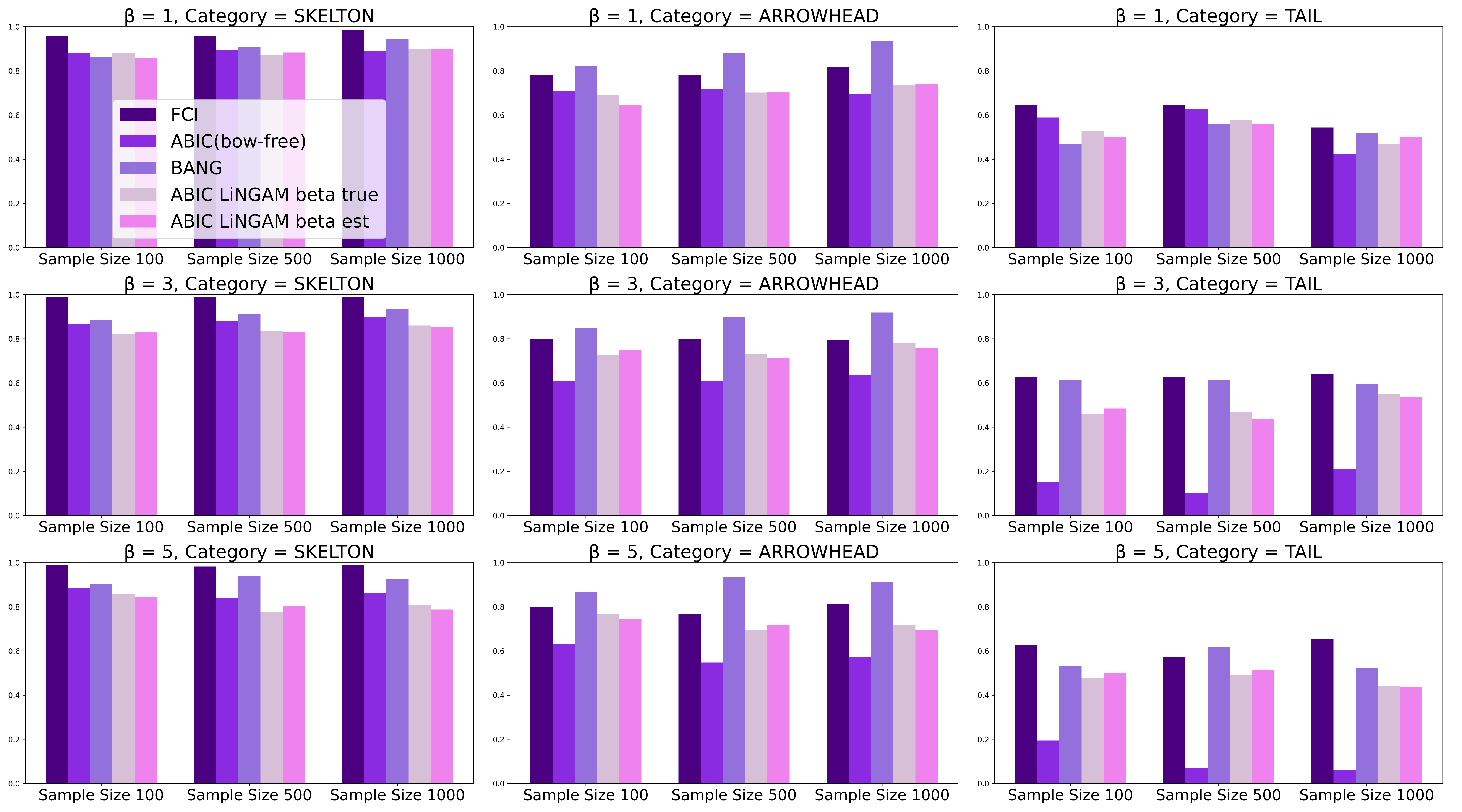}  % Include the PDF file
      \caption{The precision results for each method with five variables.}
      \label{fig:precision_5_vars}
      \vspace{1em}
  \end{figure}
  
  \begin{figure}[htbp]
    \vspace{-10em}
      \centering
      \includegraphics[bb=0 0 2200 1600, width=\textwidth]{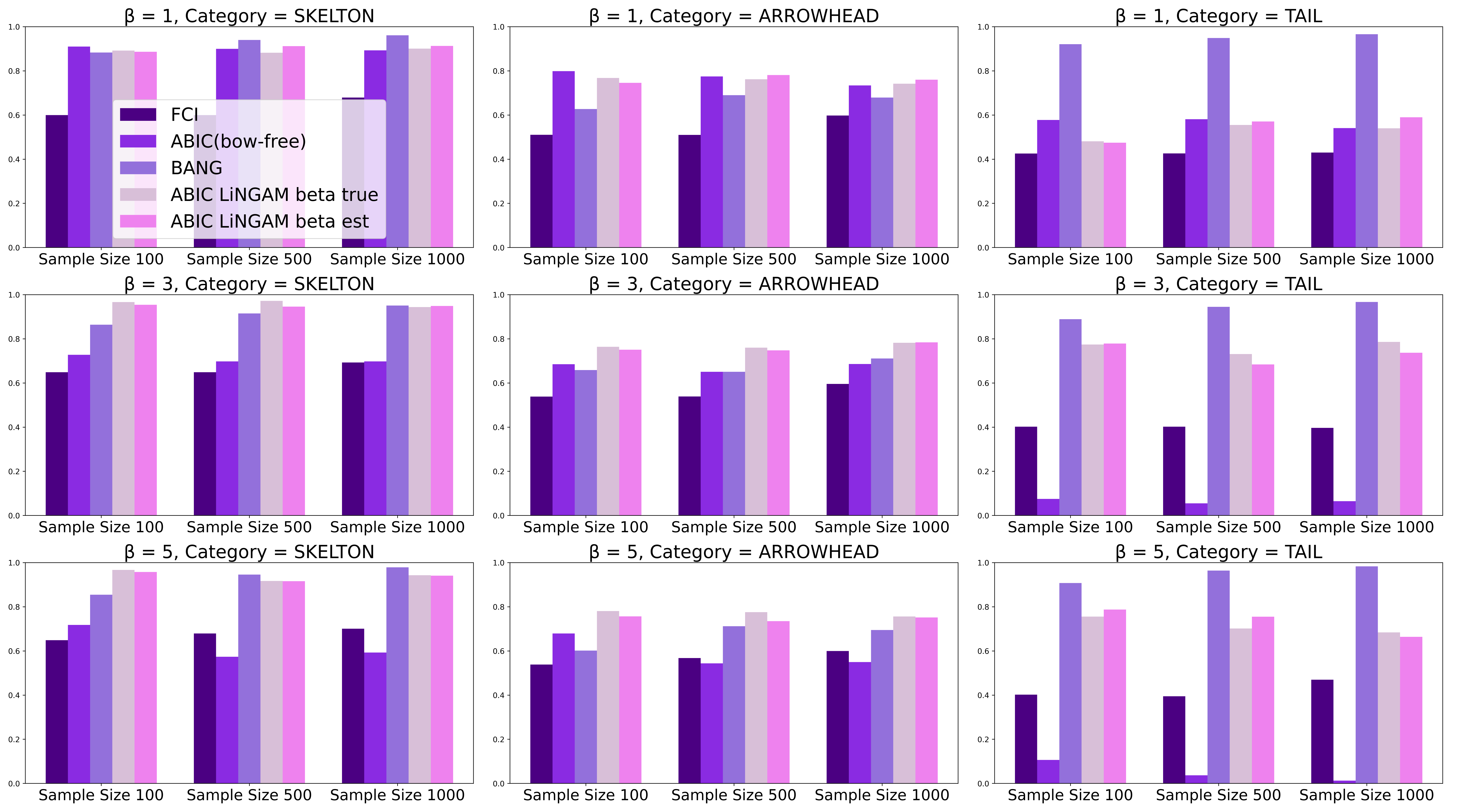}  % Include the PDF file
      \caption{The recall results for each method with five variables.}
      \label{fig:recall_5_vars}
      \vspace{1em}
  \end{figure}
  
  \begin{figure}[htbp]
    \vspace{-10em}
      \centering
      \includegraphics[bb=0 0 2200 1600, width=\textwidth]{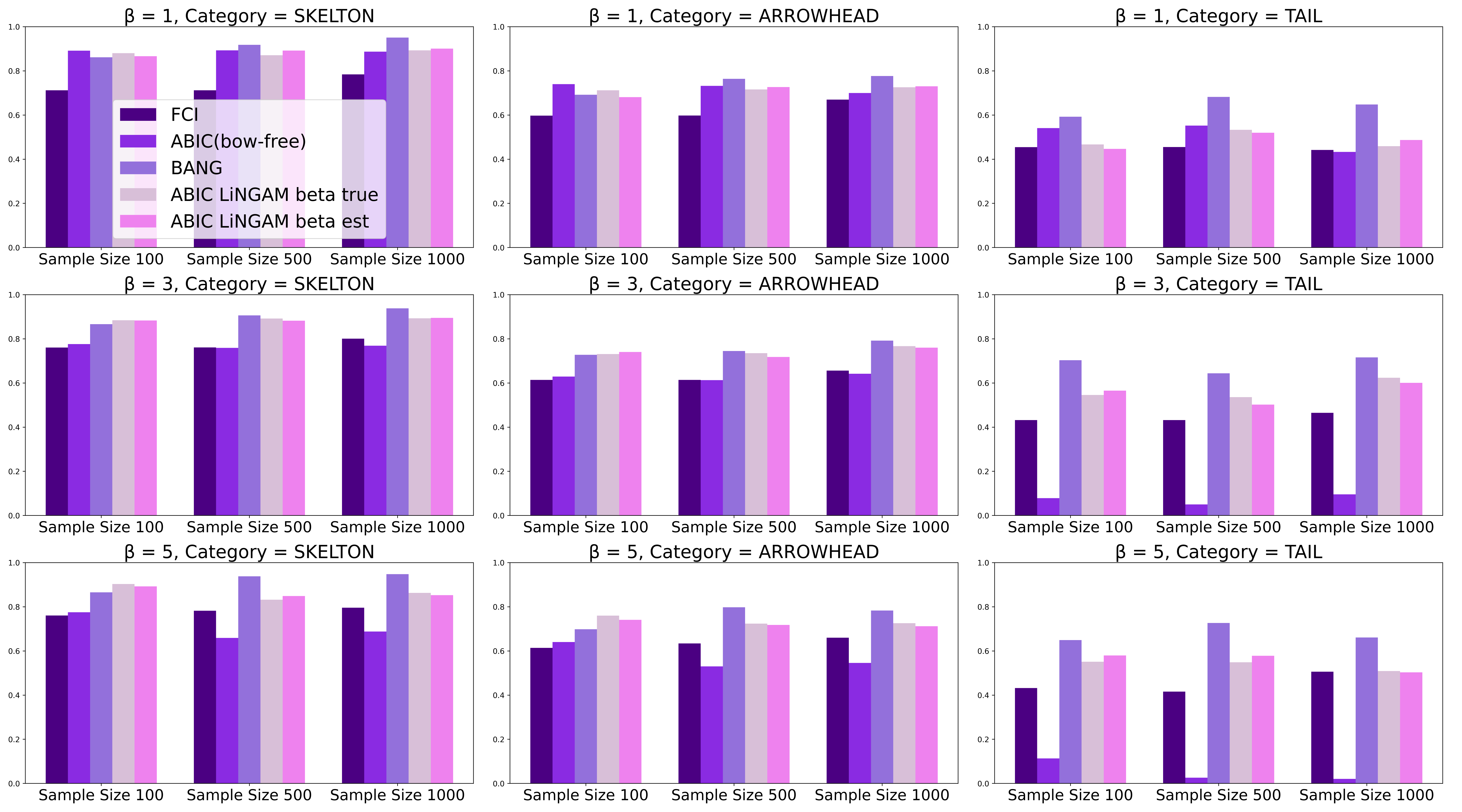}  % Include the PDF file
      \caption{The F1-score results for each method with five variables.}
      \label{fig:f1_5_vars}
      \vspace{1em}
  \end{figure}
  
  \begin{figure}[htbp]
    \vspace{-10em}
      \centering
      \includegraphics[bb=0 0 2200 1600, width=\textwidth]{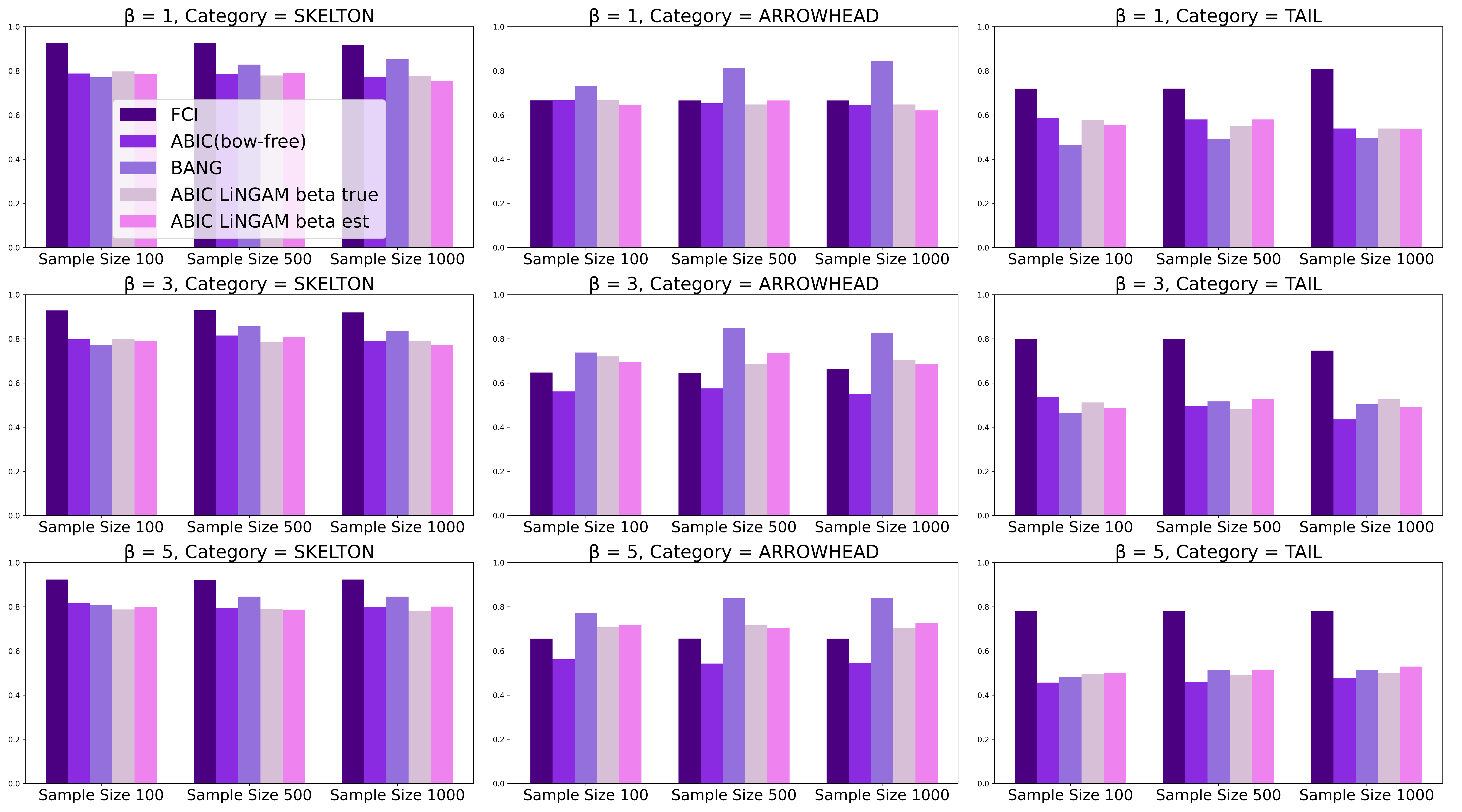}  % Include the PDF file
      \caption{The precision results for each method with ten variables.}
      \label{fig:precision_10_vars}
      \vspace{1em}
  \end{figure}
  
  \begin{figure}[htbp]
    \vspace{-10em}
      \centering
      \includegraphics[bb=0 0 2200 1600, width=\textwidth]{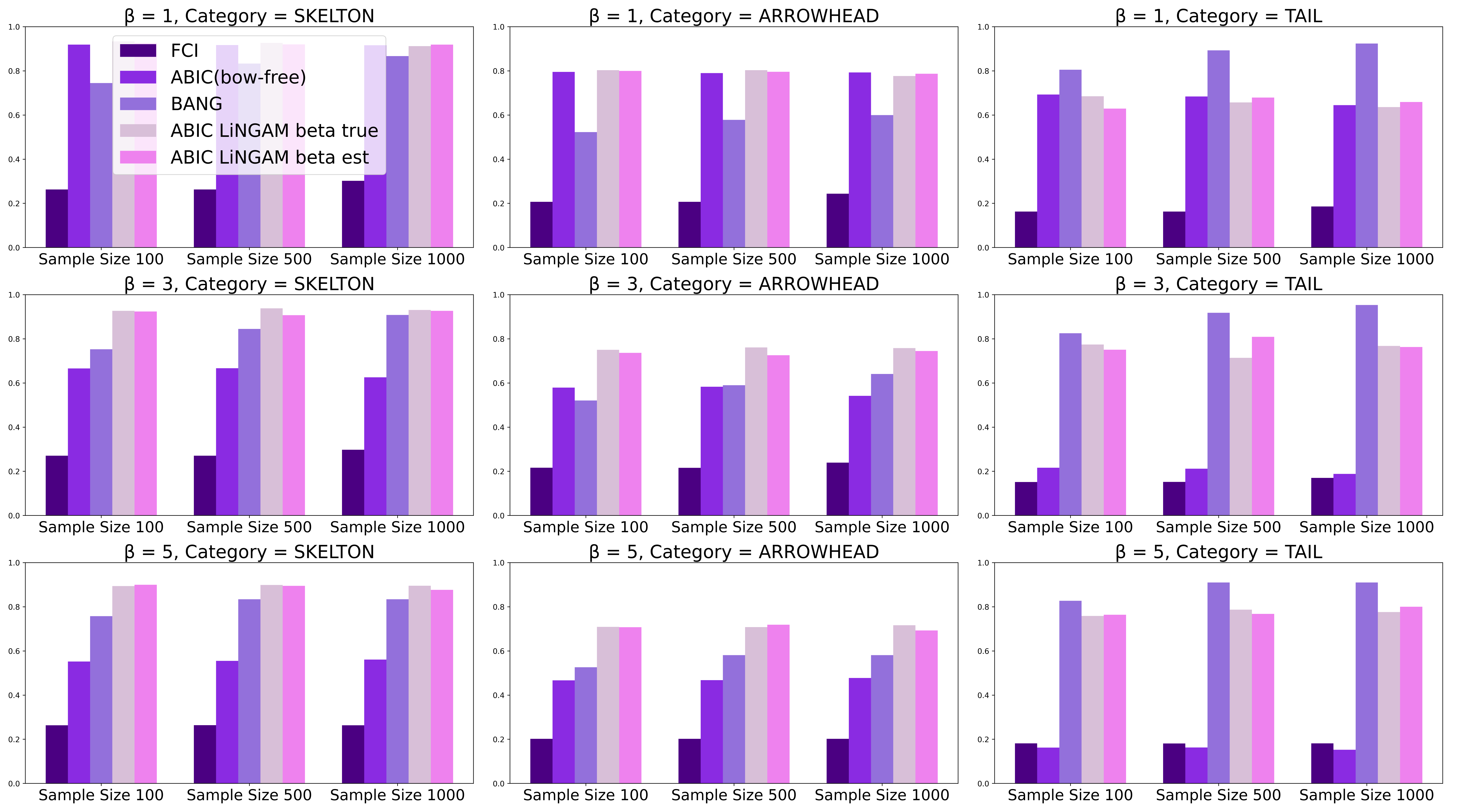}  % Include the PDF file
      \caption{The recall results for each method with ten variables.}
      \label{fig:recall_10_vars}
      \vspace{1em}
  \end{figure}
  
  \begin{figure}[htbp]
    \vspace{-10em}
      \centering
      \includegraphics[bb=0 0 2200 1600, width=\textwidth]{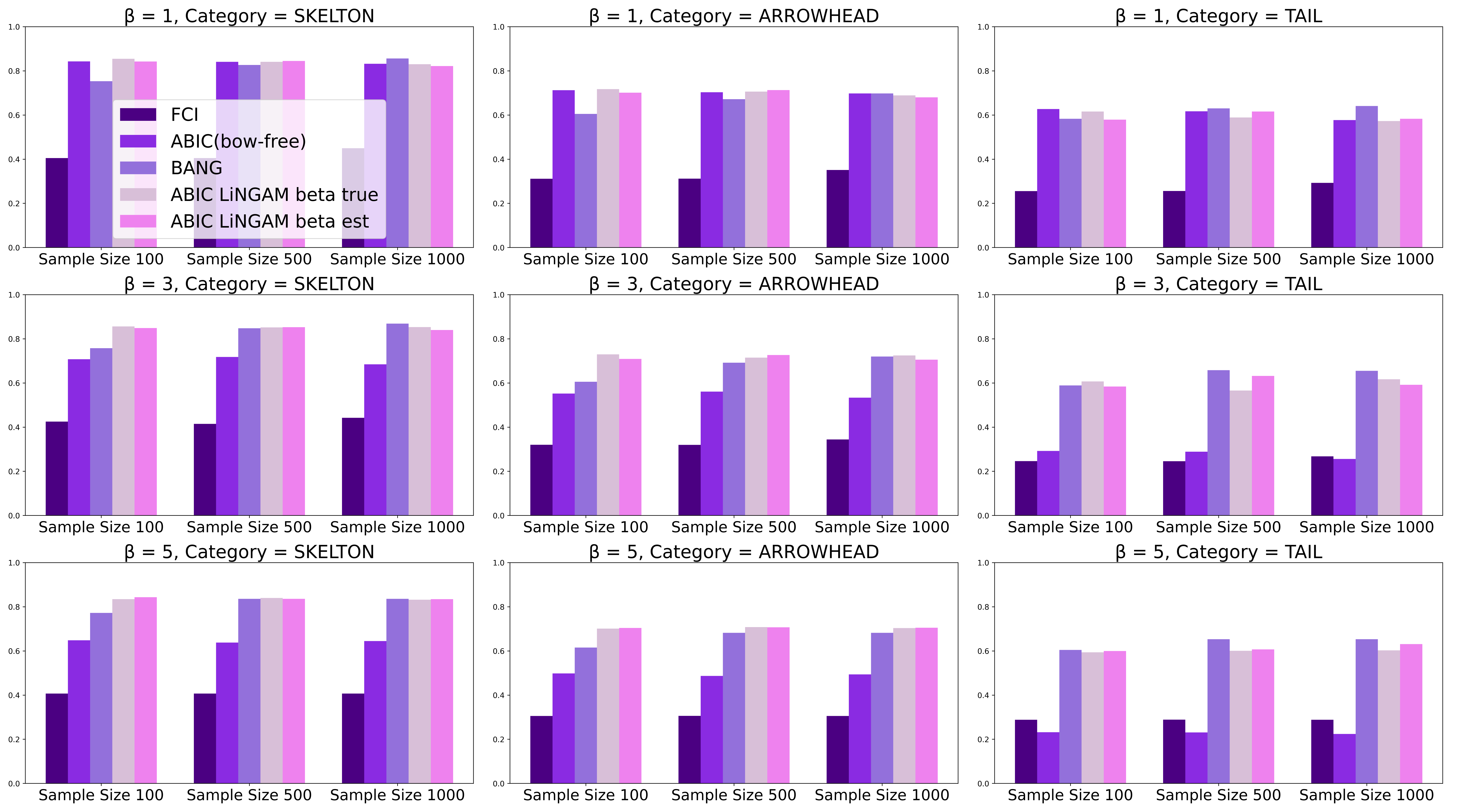}  % Include the PDF file
      \caption{The F1-score results for each method with ten variables.}
      \label{fig:f1_10_vars}
      \vspace{1em}
  \end{figure}
  
  We evaluated the performance of the proposed method, ABIC LiNGAM, using the metrics "recall," "precision," and "F1-score," and compared it with that of existing methods such as BANG. The simulation results (see Figures 1-6) showed that ABIC LiNGAM could estimate causal structures, including directions, even in the presence of unmeasured variables.
  
  First, in the case of $\beta=1$, ABIC LiNGAM exhibited a performance comparable to that of ABIC in SKELETON, accurately estimating structures without considering directions. This is consistent with the fact that ABIC can estimate up to the Markov equivalence class in linear Gaussian systems. When \( \beta = 1 \), the distribution becomes normal; thus,  ABIC LiNGAM cannot estimate the direction.
  
  Further, in the non-Gaussian distribution cases with $\beta \neq 1$, ABIC LiNGAM showed high accuracy in ARROWHEAD and TAIL, confirming that it can estimate directions. For example, under the conditions of five variables, a sample size of 500, and $\beta=3$, the recall of ARROWHEAD for ABIC LiNGAM (beta true) was 0.760, precision was 0.733, and F1-score was 0.735, which outperformed BANG's recall of 0.651, precision of 0.898, and F1-score of 0.745. Additionally, ABIC's TAIL had a recall of 0.055, a precision of 0.103, and F1-score of 0.050, indicating that it could hardly infer directions as it assumes a normal distribution. We also confirmed that the estimation accuracy remained stable even when $\beta$ increased. Specifically, under the conditions of ten variables, a sample size of 500, and $\beta=5$, the recall of ARROWHEAD for ABIC LiNGAM (beta est) was 0.719, precision was 0.705, and F1-score was 0.707, which was not significantly different from those under $\beta=3$.
  
  Moreover, it was confirmed that the accuracy of ABIC LiNGAM does not change significantly, even when the number of variables increased. When the number of variables was ten, for a sample size of 500 and $\beta=3$, the recall of ARROWHEAD for ABIC LiNGAM (beta est) was 0.726, precision was 0.736, and F1-score was 0.727, similar to the case when the number of variables was five.
  
  In terms of method comparison, there was a tendency for the performance difference between ABIC LiNGAM and BANG to be narrow. When the number of variables was ten, under the conditions of a sample size of 500 and $\beta=3$, ABIC LiNGAM (beta est) showed a recall of 0.726, a precision of 0.736, and F1-score of 0.727 for ARROWHEAD, outperforming BANG's recall of 0.590 and F1-score of 0.692. This suggests that the proposed method is effective for high-dimensional data analysis.
  
  Overall, BANG showed the highest accuracy in inferring directions. However, as the number of variables increased, the accuracies of BANG and ABIC LiNGAM (beta est) became closer. For instance, when the number of variables was five, under the condition of a sample size of 500 and $\beta=3$, the recall of TAIL for ABIC LiNGAM (beta est) was 0.684, precision was 0.436, and F1-score was 0.502, while BANG's TAIL had a recall of 0.945, a precision of 0.526, and F1-score of 0.644, indicating that BANG inferred directions better than ABIC LiNGAM (beta est). However, when the number of variables was ten, the difference in accuracy between the two methods decreased.
  
  These results were based on the numerical values obtained from the simulations. In particular, since it was important to estimate the directions in this study, we focused on ARROWHEAD and TAIL. The results showed that the proposed method, ABIC LiNGAM, could estimate causal structures including directions with high accuracy, even when unmeasured variables exist and the data follow a non-Gaussian distribution, while accurately estimating SKELETON when the data follow a normal distribution. A score-based method is considered more suitable when the overall and unified quantitative evaluation criterion is used to measure the consistency between the data and the model structure, rather than relying solely on conditional independence tests to determine the model structure as in constraint-based methods. The proposed method achieved an accuracy comparable to that of the state-of-the-art BANG method.

  \subsection{Performance Evaluation on Real-world Data}
  
\begin{figure}[htbp]
      \centering
      \includegraphics[bb=0 0 720 480, width=\textwidth]{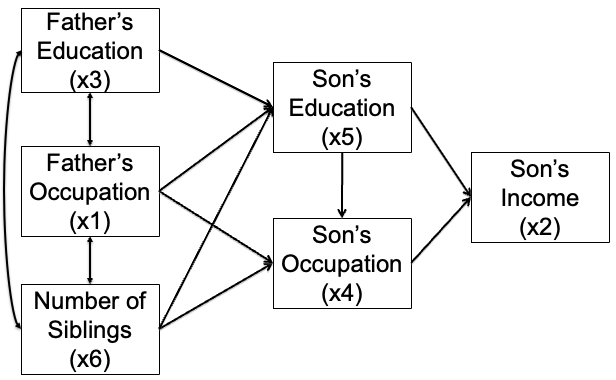}
      \caption{Variables and causal relations in the General Social Survey dataset used for the evaluation.}
      \label{fig:bollen_true_graph}
      \vspace{-1em}
  \end{figure}

  We analyzed the General Social Survey dataset obtained from a sociological data repository (https://gss.norc.org/). This dataset was also used in the evaluation of DirectLiNGAM by Shimizu et al. (2011). The sample size was 1380. The variables and their possible directions are shown in Figure \ref{fig:bollen_true_graph}. These directions were determined based on domain knowledge and time order from \cite{duncan1972occupational}.
  
  For performance evaluation, we compared the methods in the same manner as in the simulations. Specifically, to compare with our proposed method, bow-free ABIC LiNGAM, we considered bow-free ABIC (\cite{bhattacharya2021differentiable}), which was the basis of our research; FCI (\cite{spirtes2000causation}), a constraint-based method that can estimate causal structures in the presence of unmeasured variables; BANG (\cite{BANG}), which can identify bow-free ADMGs (at the 5\% significance level ). In this case, since we estimated the causal structures by assuming that the error terms of the data followed a multivariate generalized normal distribution, we inferred the shape parameter $\beta$ of the multivariate generalized normal distribution, as in the simulation, and estimated the causal structures using the inferred parameter. In addition, since estimating $\beta$ using the whole dataset resulted in an estimate less than 1 and unstable estimation when using that parameter(the estimation itself was feasible, but \( \delta \) had all elements equal to 0, and \( \Omega \) had only diagonal elements equal to 1), we estimated $\beta$ using each variable individually and used the highest $\beta$ value for causal structure identification.
  
  \begin{figure}[htbp]
      \centering
      \label{fig:bollen_results}
      \includegraphics[bb=0 0 720 480, width=\textwidth]{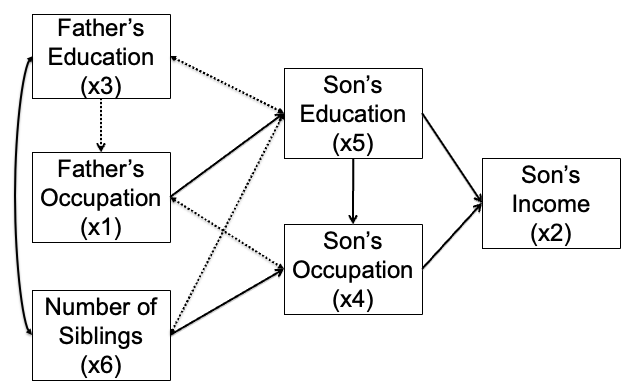}  % Include the PDF file
      \caption{Causal graph produced by ABIC LiNGAM: The dashed lines represent predicted arrows that differ from the true arrows.}
  \end{figure}
  
  \begin{table*}[htbp]
      \caption{Experimental Results on Bollen Data}
      \label{tab:bollen_results}
      \setlength{\tabcolsep}{2pt} % 列間の余白を縮める
      \renewcommand{\arraystretch}{0.9} % 行間を縮める
  \begin{tabular}{|l|ccc|ccc|ccc|}
      \hline
      \textbf{Method} & \multicolumn{3}{c|}{\textbf{SKELETON}} & \multicolumn{3}{c|}{\textbf{ARROWHEAD}} & \multicolumn{3}{c|}{\textbf{TAIL}} \\
      \cline{2-10}
      & Recall & Precision & F1-score & Recall & Precision & F1-score & Recall & Precision & F1-score \\
      \hline
      FCI & 0.727 & 1.000 & 0.842 & 0.643 & 0.692 & 0.666 & 0.250 & 0.667 & 0.363 \\
      BANG & 0.727 & 0.889 & 0.800 & 0.643 & 0.643 & 0.643 & 0.250 & 0.500 & 0.333 \\
      ABIC & 0.455 & 1.000 & 0.625 & 0.429 & 0.600 & 0.500 & 0.000 & 0.000 & 0.000 \\
      \makecell{ABIC LiNGAM \\ beta est} & 0.818 & 1.000 & 0.900 & 0.714 & 0.800 & 0.740 & 0.500 & 0.800 & 0.615 \\
      \hline
  \end{tabular}
  \end{table*}
  
  The proposed method, ABIC LiNGAM beta est, showed superior performance in SKELETON compared with the other methods, with a recall of 0.818, a precision of 1.000, and F1-score of 0.900 (see Table 2). This indicates that the proposed method can estimate SKELETON with higher accuracy than ABIC. While ABIC had a recall of 0.455, a precision of 1.000, and F1-score of 0.625, the proposed method was superior in terms of both recall and F1-score.
  
  Furthermore, the proposed method showed excellent performance for ARROWHEAD and TAIL, which considered the causal directions. In ARROWHEAD, the recall was 0.714, precision was 0.800, and F1-score was 0.740, surpassing BANG's recall of 0.643, precision of 0.643, and F1-score of 0.643. In TAIL, the proposed method had a recall of 0.500, a precision of 0.800, and F1-score of 0.615, which were higher than BANG's recall (0.250), precision (0.500), and F1-score (0.333). Furthermore, as shown in Figure 9, the true causal structure is estimated to some extent. These results demonstrate that the proposed method can estimate the causal structures, including directions, more accurately.
  
  \subsection{Performance Evaluation on Real Data Using Prior Knowledge}
  In ABIC LiNGAM, prior knowledge can be incorporated into the inference process following the code implementation of ABIC available at https://gitlab.com/rbhatta8/dcd. Specifically, by defining hierarchical causal orders (tiers) and incorporating prior knowledge that certain variables are unconfounded, we can restrict the parameter ranges (bounds) considered during the estimation. Consequently, edges that contradict the causal order, as well as bidirectional edges among unconfounded variables, are represented with predetermined constraints in the parameter space (e.g., fixed to zero) and are thus automatically excluded during the inference process. Thereby, explicitly reflecting prior knowledge in the model ensures that assumptions regarding causal directionality and the absence of confounding factors are maintained, enabling an efficient structural estimation.
  
  For example, the General Social Survey dataset obtained from a sociological data repository (https://gss.norc.org/) includes variables pertaining to parents and children. If a causal path exists from the parents to the children, a causal path from the children to the parents is not possible. In the following section, we consider the inferences with a hierarchical causal structure. Specifically, we divided the variables into two groups with a two-tiered structure—{Father’s Occupation, Father’s Education} and {Number of Siblings, Son’s Education, Son’s Occupation, Son’s Income}—and prohibited the existence of directed edges from children’s variables to parents’ variables.
  
  \begin{figure}[htbp]
      \centering
      \label{fig:bollen_results_prior}
      \includegraphics[bb=0 0 720 480, width=\textwidth]{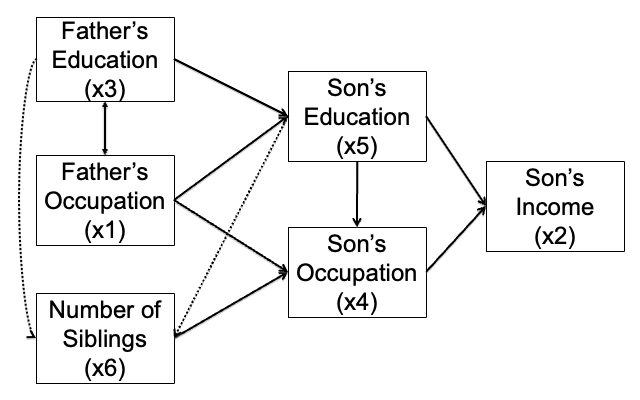}  % Include the PDF file
      \caption{Causal graph produced by ABIC LiNGAM incorporating prior knowledge: The dashed lines represent predicted arrows that differ from the true arrows.}
  \end{figure}
  
  \begin{table*}[htbp]
      \caption{Experimental Results for Bollen Data incorporating prior knowledge}
      \label{tab:bollen_results}
      \setlength{\tabcolsep}{2pt} % 列間の余白を縮める
      \renewcommand{\arraystretch}{0.9} % 行間を縮める
  \begin{tabular}{|l|ccc|ccc|ccc|}
      \hline
      \textbf{Method} & \multicolumn{3}{c|}{\textbf{SKELETON}} & \multicolumn{3}{c|}{\textbf{ARROWHEAD}} & \multicolumn{3}{c|}{\textbf{TAIL}} \\
      \cline{2-10}
      & Recall & Precision & F1-score & Recall & Precision & F1-score & Recall & Precision & F1-score \\
      \hline
      \makecell{ABIC LiNGAM \\ beta est} & 0.818 & 1.000 & 0.900 & 0.714 & 0.800 & 0.740 & 0.500 & 0.800 & 0.615 \\
      \makecell{ABIC LiNGAM \\ beta est \\ prior knowledge} & 0.818 & 1.000 & 0.900 & 0.692 & 0.900 & 0.782 & 0.857 & 0.750 & 0.800 \\
      \hline
  \end{tabular}
  \end{table*}
  
  As demonstrated by the results shown in Figure 10 and Table 3, incorporating prior information into ABIC LiNGAM leads to a higher estimation accuracy compared to the approach without such information. The estimated graph aligns closely with the presumed true causal structure. This suggests that incorporating prior knowledge into ABIC LiNGAM is feasible and that leveraging such information can yield improved accuracy in practical applications.

\section{Conclusion}
  
  In this study, based on the method proposed by \cite{bhattacharya2021differentiable}, we introduced ABIC LiNGAM, which extends the LiNGAM method for causal structure estimation in the presence of unmeasured variables. By assuming that the error terms follow a multivariate generalized normal distribution, we showed that we can estimate not only up to the Markov equivalence class but also the directions in the causal structure. Additionally, since ABIC LiNGAM can accurately estimate the SKELETON when the data follow a normal distribution, it can be considered a more generalized approach than the ABIC LiNGAM proposed in the previous study by \cite{bhattacharya2021differentiable}. We also proved the identifiability of the parameters in ADMGs when the error terms follow a multivariate generalized normal distribution. Through simulations and experiments using real-world data, we confirmed that the proposed method can estimate causal structures with an accuracy comparable to that of existing methods. The proposed method is expected to provide a useful framework for causal discovery in real-world situations with unmeasured variables. In future work, we will explore methods to reduce the estimation time, extending the proposed method to nonlinear data structures, and applying it to mixed data, including discrete variables.

\section{Acknowledgments}
  Shohei Shimizu was partially supported by JST CREST JPMJCR22D2, and Grant-in- Aid for Scientific Research (C) from the Japan Society for the Promotion of Science (JSPS) \#20K11708.


\begin{thebibliography}{}

    \bibitem{bhattacharya2021differentiable}
    R. Bhattacharya, T. Nagarajan, D. Malinsky, and I. Shpitser.
    \newblock Differentiable causal discovery under unmeasured confounding.
    \newblock In {\em International Conference on Artificial Intelligence and Statistics}, pages 2314--2322. PMLR, March 2021.
    
    \bibitem{brito2002recursive}
    C. Brito and J. Pearl.
    \newblock A new identification condition for recursive models with correlated errors.
    \newblock {\em Structural Equation Modeling}, 9(4):459--474, 2002.
    
    \bibitem{chickering1996optimal}
    D. M. Chickering.
    \newblock Learning {B}ayesian Networks is {NP}-Complete.
    \newblock In {\em Artificial Intelligence and Statistics}, pages 121--130, 1996.
    
    \bibitem{chickering2002optimal}
    D. M. Chickering.
    \newblock Optimal structure identification with greedy search.
    \newblock {\em Journal of Machine Learning Research}, 3(Nov):507--554, 2002.
    
    \bibitem{colombo2012learning}
    D. Colombo, M. H. Maathuis, M. Kalisch, and T. S. Richardson.
    \newblock Learning high-dimensional directed acyclic graphs with latent and selection variables.
    \newblock {\em The Annals of Statistics}, pages 294--321, 2012.
    
    \bibitem{duncan1972occupational}
    O. D. Duncan and D. L. Featherman.
    \newblock Psychological and cultural factors in the process of occupational achievement.
    \newblock {\em Social Science Research}, 1(2):121--145, 1972.
    
    \bibitem{dorton2009computing}
    M. Dorton, M. Eichler, and T. Richardson.
    \newblock Computing maximum likelihood estimates in recursive linear models with correlated errors.
    \newblock {\em Journal of Machine Learning Research}, 10, 2009.
    
    \bibitem{okamoto1973eigenvalues}
    M. Okamoto.
    \newblock Distinctness of the eigenvalues of a quadratic form in a multivariate sample.
    \newblock {\em The Annals of Statistics}, pages 763--765, 1973.
    
    \bibitem{gomez1998power}
    E. G{\'o}mez, M. A. Gomez-Villegas, and J. M. Mar{\'i}n.
    \newblock A multivariate generalization of the power exponential family of distributions.
    \newblock {\em Communications in Statistics-Theory and Methods}, 27(3):589--600, 1998.
    
    \bibitem{hoyer2008estimation}
    P. O. Hoyer, S. Shimizu, A. J. Kerminen, and M. Palviainen.
    \newblock Estimation of causal effects using linear {non-Gaussian} causal models with hidden variables.
    \newblock {\em International Journal of Approximate Reasoning}, 49(2):362--378, 2008.
    
    \bibitem{maeda2020rcd}
    T. N. Maeda and S. Shimizu.
    \newblock {RCD}: Repetitive causal discovery of linear {non-Gaussian} acyclic models with latent confounders.
    \newblock In {\em International Conference on Artificial Intelligence and Statistics}, pages 735--745. PMLR, June 2020.
    
    \bibitem{pearl2009causality}
    J. Pearl.
    \newblock {\em Causality}.
    \newblock Cambridge University Press, 2009.
    
    \bibitem{shimizu2006linear}
    S. Shimizu, P. O. Hoyer, A. Hyv{\"a}rinen, and A. Kerminen.
    \newblock A linear non-{Gaussian} acyclic model for causal discovery.
    \newblock {\em Journal of Machine Learning Research}, 7(10), 2006.
    
    \bibitem{shpitser2006identification}
    I. Shpitser and J. Pearl.
    \newblock Identification of joint interventional distributions in recursive semi-{Markovian} causal models.
    \newblock In {\em Proceedings of AAAI}, pages 1219--1226, July 2006.
    
    \bibitem{spirtes1991algorithm}
    P. Spirtes and C. Glymour.
    \newblock An algorithm for fast recovery of sparse causal graphs.
    \newblock {\em Social Science Computer Review}, 9(1):62--72, 1991.
    
    \bibitem{spirtes2000causation}
    P. L. Spirtes, C. N. Glymour, and R. Scheines.
    \newblock {\em Causation, Prediction, and Search}.
    \newblock MIT Press, 2000.
    
    \bibitem{Vermaconstraint}
    T. Verma and J. Pearl.
    \newblock Equivalence and synthesis of causal models.
    \newblock In {\em Proceedings of the 6th Annual Conference on Uncertainty in Artificial Intelligence}, 1990.
    
    \bibitem{zheng2018dags}
    X. Zheng, B. Aragam, P. K. Ravikumar, and E. P. Xing.
    \newblock {DAG}s with no tears: Continuous optimization for structure learning.
    \newblock {\em Advances in Neural Information Processing Systems}, 31:9472--9483, 2018.
    
    \bibitem{BANG}
    Y. Samuel Wang and Mathias Drton.
    \newblock Causal discovery of linear {non-Gaussian} causal models with unobserved confounding.
    \newblock {\em Journal of Machine Learning Research}, 21:1329--1359, 2024.
    
    \bibitem{wright1960}
    S. Wright.
    \newblock Path coefficients and path regressions: alternative or complementary concepts?
    \newblock {\em Biometrics}, 16(2):189--202, 1960.
    
\end{thebibliography}

\appendix
\section{Proof of the Identifiability of Parameters in Bow-Free ADMGs with Multivariate Generalized Normal Distributions}

\textbf{Theorem 1}

Let $G$ be a bow-free ADMGs with error terms following a multivariate generalized normal distribution, and let the set of parameters of $G$ be $\theta = \{\delta, \Omega\}$. Then, for almost all $\theta$, the following holds:
\[
\Sigma(\theta) = \Sigma(\theta')
\]
implies $\theta = \theta'$.

In other words, if two parameter sets $\theta$ and $\theta'$ give the same covariance matrix $\Sigma$, then $\theta$ and $\theta'$ must be identical, except possibly when $\theta$ belongs to a set of Lebesgue measure zero. If the two lemmas described later can be proven for the case where the error terms follow a multivariate generalized normal distribution, this theorem can be demonstrated using the same proof as in \cite{brito2002recursive}.

\textbf{Definition 1}

A \textit{path} in a graph is a sequence of edges (directed or bidirectional), where each edge starts from the node where the previous edge ends. A \textit{directed path} consists only of directed edges all pointing in the same direction. A node $X$ is called an \textit{ancestor} of a node $Y$ if there is a directed path from $X$ to $Y$. A path is said to be \textit{blocked} if there is a node $Z$ on the path such that there are consecutive edges pointing toward $Z$ (e.g., $\dots \rightarrow Z \leftarrow \dots$). In this case, $Z$ is called a \textit{collider}.

\textbf{Definition 2}

In a DAG, the \textit{depth} of a node is defined as the length (number of edges) of the longest path directed from its ancestors to that node.

\textbf{Lemma 1}

Let $X$ and $Y$ be the nodes in a bow-free ADMG with $depth(X) \geq depth(Y)$. Then, all paths between $X$ and $Y$ that include a node $Z$ satisfying $depth(Z) \geq depth(X)$ are blocked by colliders. This lemma is based on graph theory and does not depend on the distribution of the error terms. It is quoted from \cite{brito2002recursive}.

\textbf{Definition 3}

For each node $Y$, the set of edges directed to $Y$, denoted by $I(Y)$, is defined as the union of the following two sets:
(a) the set of all directed edges pointing to $Y$,
(b) the set of all bidirectional edges between $X$ and $Y$, where $depth(X) < depth(Y)$.

\textbf{Lemma 2}

Let \( Y \) be a variable at depth \( k \) in a bow-free ADMG. Assume that the parameters of all edges connecting variables of a depth less than \( k \) are identifiable. Then, in almost all cases, the parameters of each edge in the set \( I(Y) \) are identifiable.

\textbf{Proof}

In \cite{brito2002recursive}, the identifiability of bow-free models was established under the assumption that the error terms follow a multivariate normal distribution. In this study, we extend this identifiability result to the case where the error terms follow the aforementioned multivariate generalized normal distribution. The proof itself draws heavily from \cite{brito2002recursive}.

Wright's method \cite{wright1960} relies on linear relationships between variables and covariance structures. Since the multivariate generalized normal distribution is closed under linear transformations(\cite{gomez1998power}), Wright's method is applicable beyond the normal distribution as long as the necessary linear conditions are satisfied. Indeed, \cite{wright1960} also mentions that Wright's method can be applied to distributions other than the normal distribution.

Let \( X = \{X_1, X_2, \dots, X_m\} \) be the set of variables with a depth less than \( k \), and suppose that these variables are connected to \( Y \) by directed or undirected edges. By the properties of bow-free ADMGs, a one-to-one correspondence exists between each variable in \( X \) and the edges in \( I(Y) \). Therefore, \( I(Y) \) can be expressed as
\[
I(Y) = \{(X_1, Y), (X_2, Y), \dots, (X_m, Y)\}.
\]

Applying Wright's method to each pair \( (X_i, Y) \) yields the following equations:
\[
\sigma_{X_i Y} = \sum_{p_i} T(p_i), \quad i = 1, \dots, m
\]
where \( \sigma_{X_i Y} \) represents the covariance between \( X_i \) and \( Y \), the sum is over all paths \( p_i \) between \( X_i \) and \( Y \) that have direct or indirect effects or associations, and \( T(p_i) \) represents the product of parameters along the path \( p_i \).

For each \( i \), let \( \lambda_i \) be the parameter corresponding to the edge \( (X_i, Y) \). The equation can be rewritten as
\[
\sigma_{X_i Y} = \lambda_i + \sum_{j \neq i} \lambda_j a_{ij}, \quad i = 1, \dots, m
\]
where the coefficients \( a_{ij} \) are functions of identifiable parameters corresponding to edges connecting variables of a depth less than \( k \). These coefficients reflect contributions from direct or indirect effects or associations involving known parameters, excluding the direct edge \( (X_i, Y) \).

Under the assumption, by the induction hypothesis, that all parameters of edges connecting variables of a depth less than \( k \) are identifiable, the coefficients \( a_{ij} \) are known quantities. Therefore, we obtain a system of \( m \) linear equations with \( m \) unknowns \( \lambda_1, \dots, \lambda_m \), which can be written in a matrix form as

\begin{equation}
    \sigma = A \lambda
\end{equation}

where
\[
\sigma = 
\begin{pmatrix}
\sigma_{X_1 Y} \\
\sigma_{X_2 Y} \\
\vdots \\
\sigma_{X_m Y}
\end{pmatrix},
\quad
\lambda = 
\begin{pmatrix}
\lambda_1 \\
\lambda_2 \\
\vdots \\
\lambda_m
\end{pmatrix},
\quad
A = 
\begin{pmatrix}
1 & a_{12} & \dots & a_{1m} \\
a_{21} & 1 & \dots & a_{2m} \\
\vdots & \vdots & \ddots & \vdots \\
a_{m1} & a_{m2} & \dots & 1
\end{pmatrix}.
\]

To establish the identifiability of the parameters \( \lambda_i \), it suffices to show that matrix \( A \) is invertible in almost all cases, that is, \( \det(A) \neq 0 \) except on a set of measure zero, considering that the left-hand side \( \sigma \) is observable. The matrix \( A \) has all diagonal elements equal to 1, and off-diagonal elements depending on the model parameters. The determinant can be expressed in terms of the diagonal and off-diagonal elements, as shown in (\ref{det A}).

\begin{equation}
    \label{det A}
    \det(A) = 1 + T,
\end{equation}

where \( T \) is either zero or a polynomial in the model parameters that do not contain any constant term.

According to a well-known result in algebraic geometry \cite{okamoto1973eigenvalues}, the set of parameter values where \( \det(A) = 0 \) has Lebesgue measure zero in the parameter space. This is because \( \det(A) = 0 \) defines an algebraic variety of a lower dimension within the parameter space. Therefore, the matrix \( A \) is invertible in almost all cases, and the system of linear equations has a unique solution.

Thus, under the given assumptions, each parameter \( \lambda_i \) is identifiable in almost all cases.

\hfill \(\square\)

\section{Hölder's Inequality}

Hölder's inequality is a fundamental result in analysis that provides estimates for sequences (or more generally, measurable functions on a measure space \((\Omega, \mu)\)) in terms of their \(L^p\)-norms. Specifically, for \(p, q \geq 1\) satisfying \(\frac{1}{p} + \frac{1}{q} = 1\), Hölder's inequality states that for any two sequences \((a_k)\) and \((b_k)\):

\begin{equation}
\sum_{k=1}^\infty |a_k b_k| \leq \left( \sum_{k=1}^\infty |a_k|^p \right)^{1/p} \left( \sum_{k=1}^\infty |b_k|^q \right)^{1/q}.
\end{equation}

Furthermore, by taking \(b_k = 1\), we obtain a useful inequality for finite sums as follows:

\begin{equation}
\left( \sum_{k=1}^n |a_k| \right)^p \leq n^{p-1} \sum_{k=1}^n |a_k|^p.
\end{equation}

This special case reflects how the \(L^p\)-norm behaves in a finite setting and is central to understanding the interplay between norms and summation.

\medskip

In our specific problem, we use Hölder's inequality to handle the terms involving \(\beta\)th powers of squared residuals. Considering the log-likelihood expression after rearrangement,

\begin{equation}
    \label{mggd organized decomposition log density4}
    \begin{aligned}
    \ell(\mu, \Sigma, \beta | X) \\  
     &= -\frac{N}{2} \log \Omega_{ii. -i} 
     -  \frac{1}{2\Omega_{ii. -i}^\beta}  \sum_{l=1}^{N} \left( \left( X_i^{(l)} - \sum_{j \in \mathrm{pa}(i)} \delta_{i,j} X_j^{(l)} - \sum_{k \in sp(i)} \Omega_{i,k} Z_k^{(l)}  \right)^2 \right)^\beta \\
     &= -\frac{N}{2} \log \Omega_{ii. -i} 
     -  \frac{1}{2\Omega_{ii. -i}^\beta} \frac{N^{\beta-1}}{N^{\beta-1}} \sum_{l=1}^{N} \left( \left( X_i^{(l)} - \sum_{j \in \mathrm{pa}(i)} \delta_{i,j} X_j^{(l)} - \sum_{k \in sp(i)} \Omega_{i,k} Z_k^{(l)}  \right)^2 \right)^\beta.
    \end{aligned}
\end{equation}

We introduce the factor \(\frac{N^{\beta-1}}{N^{\beta-1}}\) to rewrite the sum in a form amenable to Hölder's inequality. Define the sequence as
\[
a_l = \left| \left( X_i^{(l)} - \sum_{j \in \mathrm{pa}(i)} \delta_{i,j} X_j^{(l)} - \sum_{k \in sp(i)} \Omega_{i,k} Z_k^{(l)} \right)^2 \right|^\beta,
\]
and let \(b_l = 1\). If we choose \(p = \beta\), and hence \(q = \frac{\beta}{\beta - 1}\) (so that \(\frac{1}{p} + \frac{1}{q} = 1\)), Hölder's inequality gives us

\[
\sum_{l=1}^{N} |a_l b_l| \leq \left(\sum_{l=1}^{N} |a_l|^\frac{\beta}{\beta}\right)^{\frac{1}{\beta}} 
\left(\sum_{l=1}^{N} |b_l|^\frac{\beta}{\beta-1}\right)^{\frac{\beta-1}{\beta}}
= \left(\sum_{l=1}^{N} a_l\right)^{\frac{1}{\beta}} N^{\frac{\beta-1}{\beta}}.
\]

Rearranging this inequality, we obtain a lower bound on \(\sum_{l=1}^N a_l\) in terms of \(N^{\beta-1}\) and the \(L^\beta\)-norm of the residuals

\[
\sum_{l=1}^{N} \left( \left( X_i^{(l)} - \cdots \right)^2 \right)^\beta 
\geq \frac{\left( \sum_{l=1}^N |X_i^{(l)} - \cdots| \right)^\beta}{N^{\beta -1}}.
\]

Substituting this bound back into the expression for \(\ell(\mu, \Sigma, \beta | X)\), we have

\begin{equation}
    \ell(\mu, \Sigma, \beta | X) 
     \geq -\frac{N}{2} \log \Omega_{ii. -i} 
     -  \frac{1}{2\Omega_{ii. -i}^\beta} \frac{1}{N^{\beta - 1}} 
     \left\| X_i - \sum_{j \in \mathrm{pa}(i)} \delta_{i,j} X_j - \sum_{k \in sp(i)} \Omega_{i,k} Z_k \right\|^{2\beta}.
\end{equation}

Thereby, Hölder's inequality is employed to provide a nontrivial lower bound on the log-likelihood by relating sums of \(\beta\)th powers of squared terms to the \(\beta\)th power of their \(L^1\)-norm, scaled appropriately by \(N^{\beta - 1}\). This facilitates a more tractable analysis of the growth behavior and bounding properties of the likelihood function.

\end{document}